\documentclass[aps,prd,reprint,superscriptaddress,nofootinbib,floatfix]{revtex4-2}
\usepackage{amsmath,amssymb,graphicx,booktabs}
\usepackage[colorlinks=true,linkcolor=blue,citecolor=blue,urlcolor=blue]{hyperref}

\newcommand{\Vus}{|V_{us}|}
\newcommand{\phifx}{\phi_{\rm FX}}
\newcommand{\e}{\varepsilon}

\usepackage{tikz}
\usepackage{pgfplots}
\pgfplotsset{compat=1.16}
\usetikzlibrary{shapes.geometric,arrows.meta,backgrounds}
\definecolor{cblue}{HTML}{2166AC}
\definecolor{cred}{HTML}{B2182B}
\definecolor{cband}{HTML}{9ECAE1}
\definecolor{cbline}{HTML}{3182BD}
\definecolor{cgreen}{HTML}{4DAF4A}
\definecolor{cboxfill}{HTML}{EAF2FA}

\begin{document}

\title{Commensurate Structure of Quark and Lepton Mixing:\\
Eighteenth Powers of One Parameter and a Digital-Clock Quantization of the Unitarity Triangles}

\author{Vernon Barger}
\affiliation{Department of Physics, University of Wisconsin--Madison, Madison, WI 53706, USA}

\date{August 10, 2026}

\begin{abstract}
We identify two commensurate structures in the measured quark and lepton mixing data, one multiplicative and one angular, and their exact relation. With the hierarchy parameter $\e\equiv|V_{ub}|^{3/10}=0.1869\pm0.0018$, the measured Fritzsch--Xing (FX) angles satisfy $\sin\theta_u=\e^{26/18}$, $\sin\theta_d=\e^{17/18}$, and $\sin\theta=\e^{34/18}$ with coefficients within $2.5\%$ of unity, and the phase is maximal, $\phifx=92.3^\circ\pm2.7^\circ$ against $\pi/2$. Unit coefficients reproduce the remaining moduli at the percent level and fix the derived Jarlskog invariant, $J=\e^{111/18}\sin\phifx$, matching the measured $3.08(14)\times10^{-5}$. The rephasing-invariant image of this structure is a unitarity triangle quantized on the $\pi/24$ lattice, a digital clock, $(\alpha,\beta,\gamma)=(12,3,9)\times7.5^\circ$, consistent with data at $1\sigma$ through $\alpha\simeq\phifx$ and the theorem $\tan\beta\simeq\sqrt{\e}$; degree-level $\gamma$ confronts $67.5^\circ$--$68^\circ$. For the leptons, where only the angular structure can act, its PMNS analog, formed from columns $1$ and $3$, free of Majorana phases, on the same lattice requires $\delta_{\rm CP}\simeq296^\circ$ in the upper octant or $244^\circ$ in the lower, testable at DUNE and Hyper-Kamiokande. The quark and lepton triangles share $\beta=22.5^\circ$, measured for quarks, predicted for leptons, and differ by a transfer of two lattice units, with corollary $\delta_{\rm CP}\simeq-\delta\simeq295^\circ$. The same $\e$ sets the neutrino mass ratio, $r\equiv m_2/m_3=\e^{19/18}$, and the PMNS first row follows in powers of $r$, $(r^{1/9},\,r^{1/3},\,\tfrac{\sqrt{3}}{2}r)$, both relations at $0.2\sigma$ and testable at JUNO. The analysis is confined to the parameterization level; no dynamics is assumed.
\end{abstract}

\maketitle

\section{Introduction}
\label{sec:intro}

The flavor parameters of the Standard Model are measured; they are not understood. In the long interval between measurement and understanding, a specific kind of analysis has repeatedly proven its worth: the identification, at the parameterization level, of compact empirical regularities that compress the data and sharpen the target for theory. The Wolfenstein expansion organized the CKM matrix around powers of a single small parameter \cite{Wolfenstein1983}; the Gatto--Sartori--Tonin relation tied the Cabibbo angle to light-quark mass ratios \cite{GST1968}; tribimaximal mixing \cite{HPS2002} and quark-lepton complementarity \cite{Minakata2004,Raidal2004} played the same role for the leptons. None of these proposals supplied dynamics, and none was diminished by that; their function was descriptive and predictive, and the model-building literature they generated is the measure of their success.

This paper is written deliberately in that tradition, and its scope should be stated at the outset. We identify two commensurate structures in the measured mixing data, one multiplicative and one angular, exhibit the exact relation between them, quantify the probability that each arises by chance, and state the prospective measurements that will confirm or exclude them. We make no assumption about symmetries, flavon content, mediator scales, or ultraviolet dynamics, and no statement in this paper depends on the existence of such a completion. Whether the patterns descend from dynamics is a separate question, on which we take no position; the patterns are offered as compressed descriptions of data and as falsifiable benchmarks, and Sec.~\ref{sec:discussion} lists what any dynamical origin would be required to explain.

The two structures are the following. Multiplicatively, the angles of the Fritzsch--Xing (FX) parameterization of the CKM matrix \cite{FritzschXing1998}, each of which is separately measurable, are integer eighteenth powers of a single hierarchy parameter $\e$, itself defined internally by the smallest CKM element; the phase $\phifx$ sits at its maximal value $\pi/2$. Additively, the angles of the unitarity triangle are integer multiples of $\pi/24=7.5^\circ$. The first structure operates in the hierarchical regime and is empty for order-one quantities, as we make quantitative below; the second is rephasing-invariant and applies wherever a nondegenerate triangle exists. The quark sector realizes both at once, and the relation between them is exact and short: the triangle pattern is the angular image of the multiplicative lattice evaluated at maximal phase, through $\alpha\simeq\phifx$ and $\tan\beta\simeq\sqrt{\e}$, the half-integer power arising as the exponent difference $26/18-17/18$. The lepton sector, where mixing is large, admits only the angular structure, and there the $\pi/24$ hypothesis converts into a definite prediction for the single unmeasured quantity, the CP phase $\delta_{\rm CP}$.

Sections~\ref{sec:fx} and \ref{sec:ninths} develop the quark mixing analysis, and Sec.~\ref{sec:qmass} extends the lattice to the quark mass ratios; Sec.~\ref{sec:stats} quantifies its statistical weight with the look-elsewhere effects stated explicitly; Secs.~\ref{sec:triangle} and \ref{sec:map} present the $\pi/24$ triangle and the map between the lattices, distilled in Sec.~\ref{sec:betatheorem} into a theorem for $\beta$; Sec.~\ref{sec:pmns} treats the leptons; Sec.~\ref{sec:discussion} discusses scope, precedents, and the requirements the patterns place on any eventual dynamical explanation.

\section{The FX parameterization and its exact data inversion}
\label{sec:fx}

The FX parameterization writes the CKM matrix as
\begin{equation}
V \;=\; R_{12}(\theta_u)\,R_{23}(\theta)\,
{\rm diag}\!\left(e^{-i\phifx},\,1,\,1\right)\,R_{12}^{\dagger}(\theta_d),
\label{eq:fx}
\end{equation}
with explicit elements
\begin{equation}
V=\begin{pmatrix}
s_u s_d c + c_u c_d\,e^{-i\phifx} & s_u c_d c - c_u s_d\,e^{-i\phifx} & s_u s\\[2pt]
c_u s_d c - s_u c_d\,e^{-i\phifx} & c_u c_d c + s_u s_d\,e^{-i\phifx} & c_u s\\[2pt]
-s_d s & -c_d s & c
\end{pmatrix},
\label{eq:fxmatrix}
\end{equation}
where $s_u=\sin\theta_u$, $s_d=\sin\theta_d$, $s=\sin\theta$, and similarly for cosines. The third row and third column are real, which yields exact inversion relations,
\begin{equation}
\tan\theta_u=\left|\frac{V_{ub}}{V_{cb}}\right|,\qquad
\tan\theta_d=\left|\frac{V_{td}}{V_{ts}}\right|,\qquad
\sin\theta=\frac{|V_{cb}|}{\cos\theta_u},
\label{eq:inversion}
\end{equation}
with $\phifx$ then determined by $\Vus$ through
\begin{equation}
\cos\phifx=\frac{s_u^2 c_d^2 c^2+c_u^2 s_d^2-\Vus^2}{2\,s_u c_u\, s_d c_d\, c}\,.
\label{eq:phifx}
\end{equation}
This observability is the property that makes the FX parameterization the right instrument for the present analysis; each parameter is fixed by data, none is a convention choice that can trade against another, and any commensurate hypothesis about the parameters is therefore directly testable. The adopted inputs and the resulting parameters are collected in Table~\ref{tab:inversion}, with uncertainties propagated by Monte Carlo from uncorrelated Gaussian inputs.

\begin{table*}[t]
\centering
\begin{tabular}{lcc}
\toprule
Input & Value & Source \\
\midrule
$\Vus$ & $0.22497\pm0.00068$ & \cite{PDG2024} \\
$|V_{cb}|$ & $0.0410\pm0.0010$ & \cite{PDG2024} \\
$|V_{ub}|$ & $0.00373\pm0.00012$ & \cite{PDG2024} \\
$|V_{td}/V_{ts}|$ & $0.2085\pm0.0040$ & \cite{PDG2024} \\
\midrule
FX parameter & Value & \\
\midrule
$\theta_u$ & $5.20^\circ\pm0.23^\circ$ & \\
$\theta_d$ & $11.78^\circ\pm0.22^\circ$ & \\
$\theta$   & $2.36^\circ\pm0.06^\circ$ & \\
$\phifx$   & $92.3^\circ\pm2.7^\circ$  & \\
\bottomrule
\end{tabular}
\caption{Adopted CKM inputs and the FX parameters obtained from the exact relations of Eqs.~(\ref{eq:inversion})--(\ref{eq:phifx}). The inputs are adopted central values based on Ref.~\cite{PDG2024}; the exclusive-leaning $|V_{ub}|$ choice is discussed in Sec.~\ref{sec:ninths}.}
\label{tab:inversion}
\end{table*}

As a consistency check on the inversion, the moduli-only determination of Table~\ref{tab:inversion} predicts the CP-violating observables with no further input; Eq.~(\ref{eq:fxmatrix}) gives $J=3.05\times10^{-5}$ against the measured $(3.08\pm0.14)\times10^{-5}$, and unitarity-triangle angles $(\alpha,\beta,\gamma)=(91.2^\circ,\,23.2^\circ,\,65.6^\circ)$ against the measured $\big(84.9^{+5.1}_{-4.5},\,22.2\pm0.7,\,65.4^{+3.8}_{-4.2}\big)^\circ$ \cite{PDG2024,LHCbGamma2021}; a recent joint BESIII--LHCb determination gives $\gamma=(71.3\pm5.0)^\circ$ \cite{BESIIILHCb2026}, consistent with these values. The measured CKM Dirac phase, $\delta=65.5^\circ\pm1.5^\circ$ in the global fit \cite{PDG2024}, emerges from $\phifx\approx92^\circ$ dressed by the small angles.

\section{The eighteenths lattice}
\label{sec:ninths}

The multiplicative hypothesis is that the FX angles are integer eighteenth powers of a single parameter, $\sin\theta_i=c_i\,\e^{n_i/18}$ with $c_i$ of order unity. To keep the analysis self-contained we define $\e$ internally, anchoring it to the smallest and most hierarchy-sensitive element,
\begin{equation}
\e \;\equiv\; |V_{ub}|^{3/10} \;=\; 0.1869\pm0.0018\,,
\label{eq:epsdef}
\end{equation}
a definition whose motivation is the lattice identity $|V_{ub}|=\sin\theta_u\sin\theta=\e^{26/18}\e^{34/18}=\e^{10/3}$ derived below; the anchor uses one modulus, and everything that follows is then a parameter-free confrontation of the remaining data. The exponent $3/10$ is itself a definition, fixed so that the anchor sits at $n=60$; the anchor's zero distance in the tables is bookkeeping, not evidence, and no chance-probability credit attaches to it.

With this $\e$, the fitted exponents of the three measured FX angles are
\begin{equation}
\begin{gathered}
\frac{18\ln\sin\theta_u}{\ln\e}=25.77,\qquad
\frac{18\ln\sin\theta_d}{\ln\e}=17.05,\\
\frac{18\ln\sin\theta}{\ln\e}=34.23,
\end{gathered}
\label{eq:fitexp}
\end{equation}
with distances to the nearest integers of $0.23$, $0.05$, and $0.23$, so that
\begin{equation}
\sin\theta_u=\e^{26/18},\qquad
\sin\theta_d=\e^{17/18},\qquad
\sin\theta=\e^{34/18}
\label{eq:lattice}
\end{equation}
with order-one coefficients $(1.022,\,0.995,\,0.979)$. The integer pattern carries a parameter-free corollary; since $34=2\times17$ and $51=3\times17$, the assignments imply, at unit coefficients,
\begin{equation}
\sin\theta \;=\; \sin^2\theta_d\,, \qquad |V_{td}| \;\simeq\; \sin^3\theta_d\,,
\label{eq:powerchain}
\end{equation}
relations among measured quantities free of $\e$ and of the anchor; the data satisfy them with coefficients $0.988$ and $1.011$. The chain does not extend to the up sector as a pure power; $\sin\theta_u=\sin^{3/2}\theta_d$ would require exponent $25.5$, a half-unit below the assignment $26$ fixed by both the anchor identity $26+34=60$ and the construction $26=2\times43-60$ of Sec.~\ref{sec:qmass}, and the lattice instead predicts the defect, $\sin\theta_u=\e^{1/36}\sin^{3/2}\theta_d$; the measured ratio $\sin\theta_u/\sin^{3/2}\theta_d=0.982$ against $\e^{1/36}=0.954$ tests the relation at the $3\%$ level, the residue carried by the order-one coefficients ($1.022/0.995^{3/2}=1.030$). The fitted exponent, $25.77\pm0.47$, cannot yet separate the half-unit; the inclusive $|V_{cb}|$ determination would decide it at $2.4\sigma$, the gain reflecting both the smaller error and the partial cancellation of the $|V_{ub}|$ uncertainty between $\sin\theta_u$ and the anchor. The denominator is the minimal one the data admit. Four of the lattice assignments reduce to ninths, $26/18=13/9$, $34/18=17/9$, and, among the moduli below, $16/18=8/9$ and $60/18=10/3$; on a pure ninths lattice the fitted exponents of $\sin\theta_u$ and $\sin\theta$ sit at distances $0.12$ and $0.11$ from integers, a stronger showing on the coarser lattice. The refinement to eighteenths is forced by the odd numerators alone. The fitted exponent of $\sin\theta_d$, $17.05\pm0.20$, selects $17/18$ at $0.3\sigma$ while rejecting both ninths neighbors, $16/18$ and $18/18$, at $5.3\sigma$ and $4.8\sigma$; its product $|V_{td}|=\e^{51/18}$ and the neutrino ratio $m_2/m_3=\e^{19/18}$ of Eq.~(\ref{eq:m23}) inherit the half-step, sitting essentially mid-lattice ($0.47$ and $0.45$) in ninths. The half-integer $\sqrt{\e}$ of Sec.~\ref{sec:map}, and with it the triangle angle $\beta$, is likewise a consequence of the single odd numerator $17$; a ninths lattice would instead require $\sin\theta_d=\e^{8/9}$, excluded at $5\sigma$, and would move $\beta$ to $21.5^\circ$.

The fourth parameter is angular and takes the maximal value; $\phifx=92.3^\circ\pm2.7^\circ$ is consistent with $\pi/2$ at the $0.9\sigma$ level, and $\pi/2$ is the distinguished point of maximal interference and maximal CP violation at fixed moduli.

Setting all coefficients to unity and $\phifx=\pi/2$, Eq.~(\ref{eq:fxmatrix}) becomes a CKM matrix with no free parameter beyond the anchor of Eq.~(\ref{eq:epsdef}). Its output is compared with data in Table~\ref{tab:lattice}. The three moduli not used in the anchor are reproduced at the percent level, and the Jarlskog invariant takes the compact closed form
\begin{equation}
\begin{split}
J \;&=\; s_u c_u\, s_d c_d\, s^2 c\,\sin\phifx\\
&\simeq\; \e^{111/18}\sin\phifx \;=\;3.2\times10^{-5}\,\sin\phifx\,,
\end{split}
\label{eq:J}
\end{equation}
with $111=26+17+2\times34$; the cosine factors reduce the full-matrix value to the $3.14\times10^{-5}$ of Table~\ref{tab:lattice}. The Cabibbo modulus illustrates how the lattice operates through interference; at $\phifx=\pi/2$ the two $1$-$2$ rotations add in quadrature,
\begin{equation}
\Vus \;\simeq\; \e^{17/18}\sqrt{1+\e}\;,
\label{eq:quad}
\end{equation}
and $\sqrt{1+\e}=1.090$ agrees with $\e^{-1/18}=1.098$ to $0.7\%$, so the effective Cabibbo exponent is shifted by one lattice unit, from $17/18$ to $16/18=8/9$, with $\e^{8/9}=0.225$ against the measured $0.22497$.

The same structure can be displayed at the level of the individual matrix elements. Table~\ref{tab:offdiag} places all six off-diagonal CKM moduli on the lattice; they are generated by four exponents, $(16,\,34,\,51,\,60)$, each of which is FX arithmetic built from the three angle integers of Eq.~(\ref{eq:lattice}): $34$ is the $2$-$3$ angle itself, $51=17+34$ and $60=26+34$ are the third-row and first-row products $|V_{td}|=s_d\,s$ and $|V_{ub}|=s_u\,s$, and $16=17-1$ is the interference shift of Eq.~(\ref{eq:quad}). We emphasize that this table adds no statistical content beyond the inputs of Table~\ref{tab:inversion}; the six moduli are tied to the four FX parameters by unitarity and the exact identities of Eq.~(\ref{eq:fxmatrix}), so the table is the same information reorganized, and no additional chance-probability credit is claimed for it in Sec.~\ref{sec:stats}. Its content is the visible coherence, the same four integers generating every placement, and the localization of the residue; the $2$-$3$ pair $(|V_{cb}|,|V_{ts}|)$ carries a single common coefficient of $0.975$--$0.977$, the $\sin\theta$ deficit already present in Eq.~(\ref{eq:lattice}), and it is the only entry beyond half a percent.

Because the pair is tied together by the FX identities, its placement is a single statement, and that statement is hostage to the unresolved tension between the inclusive and exclusive determinations of $|V_{cb}|$; the adopted $0.0410\pm0.0010$ sits between them. On the inclusive determination, $|V_{cb}|=(42.16\pm0.51)\times10^{-3}$ \cite{Gambino2021}, the fitted exponent is $33.98$ and the assignment $\e^{34/18}$ is satisfied essentially exactly, while the exclusive average $(39.46\pm0.53)\times10^{-3}$ \cite{FLAG2024} gives $34.69$, off the assigned integer; the experimental resolution of the tension therefore bears directly on the lattice, in either direction.

The anchor choice is also not delicate; anchoring instead to the Cabibbo element through $\e=|V_{us}|^{9/8}=0.18670$ reproduces $\e=|V_{ub}|^{3/10}=0.18686$ at the $0.1\%$ ($0.09\sigma$) level, so the two natural definitions of the lattice parameter agree within current uncertainties. The anchor does, however, inherit the parallel tension in $|V_{ub}|$ itself; the adopted $0.00373\pm0.00012$ is close to the exclusive determinations, while inclusive values near $4.1\times10^{-3}$ \cite{PDG2024} would give $\e=0.193$, outside the estimator band of Table~\ref{tab:estimators} altogether and moving the $\sin\theta_d$ placement from $17.05$ to approximately $17.4$. The ten non-anchor estimators thus adjudicate the tension from within; led by $|V_{us}|^{9/8}$, they require $|V_{ub}|=\e^{10/3}\simeq3.72\times10^{-3}$, so the lattice takes the exclusive side of the $|V_{ub}|$ dispute and, like the $2$-$3$ pair above, is falsifiable by its experimental resolution.

This observation generalizes into an enumeration. Every expression whose lattice exponents total $18/18$ returns $\e$ itself, and the assignments of Table~\ref{tab:offdiag} generate a complete list of the natural forms: the six single-element powers $|V_{us}|^{9/8}$, $|V_{cd}|^{9/8}$, $|V_{cb}|^{9/17}$, $|V_{ts}|^{9/17}$, $|V_{td}|^{6/17}$, and $|V_{ub}|^{3/10}$; the power-free ratios built on the exponent difference $34-16=18$, of which $|V_{cb}|/\Vus$ is representative; the squared ratio $(|V_{ub}|/|V_{td}|)^2$, which equals $(s_u/s_d)^2$ by the FX identities and is therefore $\tan^2\beta$ at leading order, the map quantity of Sec.~\ref{sec:map} measured directly; the CP-derived form $(J/\sin\phifx)^{6/37}$; and the cross-sector forms $(m_2/m_3)^{18/19}$ and $(m_\mu/2m_\tau)^{9/19}$ that follow from Eqs.~(\ref{eq:m23}) and (\ref{eq:massmass}). Table~\ref{tab:estimators} collects them.

\begin{figure*}[t]
\centering
\resizebox{0.62\textwidth}{!}{%
\begin{tikzpicture}
\begin{axis}[width=11.6cm,height=9.6cm,
  xmin=0.176,xmax=0.206,ymin=-0.7,ymax=10.7,
  xtick={0.180,0.185,0.190,0.195,0.200,0.205},
  xticklabel style={/pgf/number format/.cd,fixed,fixed zerofill,precision=3},
  xlabel={estimator of $\varepsilon$},
  ytick={10,9,8,7,6,5,4,3,2,1,0},yticklabels={{$|V_{us}|^{9/8}$},{$|V_{cd}|^{9/8}$},{$|V_{cb}|^{9/17}$},{$|V_{ts}|^{9/17}$},{$|V_{td}|^{6/17}$},{$|V_{ub}|^{3/10}$ (anchor)},{$|V_{cb}|/|V_{us}|$},{$(|V_{ub}|/|V_{td}|)^{2}$},{$(J/\sin\phi_{\rm FX})^{6/37}$},{$(m_2/m_3)^{18/19}$},{$(m_\mu/2m_\tau)^{9/19}$}},
  ytick style={draw=none}, tick label style={font=\small},
  axis line style={line width=0.5pt}]
\fill[cband,opacity=0.55] (axis cs:0.18510,-0.7) rectangle (axis cs:0.18870,10.7);
\draw[cbline,line width=0.8pt] (axis cs:0.18690,-0.7) -- (axis cs:0.18690,10.7);
\addplot[only marks,mark=*,mark size=1.7pt,color=black,
  error bars/.cd,x dir=both,x explicit,
  error bar style={line width=0.7pt},error mark options={line width=0.7pt,mark size=1.9pt,rotate=90}]
  coordinates { (0.18670,10) +- (0.00063,0) (0.18660,9) +- (0.00063,0) (0.18433,8) +- (0.00238,0) (0.18457,7) +- (0.00171,0) (0.18641,6) +- (0.00138,0) (0.18225,4) +- (0.00448,0) (0.18943,3) +- (0.01456,0) (0.18553,2) +- (0.00137,0) (0.18862,1) +- (0.00271,0) };
\addplot[only marks,mark=diamond*,mark size=3pt,mark options={fill=white,draw=black},
  error bars/.cd,x dir=both,x explicit,
  error bar style={line width=0.7pt},error mark options={line width=0.7pt,mark size=1.9pt,rotate=90}]
  coordinates { (0.18686,5) +- (0.00180,0) };
\addplot[only marks,mark=square*,mark size=1.9pt,color=black!45,
  error bars/.cd,x dir=both,x explicit,
  error bar style={line width=2.4pt,color=black!45},error mark=none]
  coordinates { (0.18817,0) +- (0.00094,0) };
\end{axis}
\end{tikzpicture}}
\caption{The eleven estimators of $\e$ from Table~\ref{tab:estimators}, each an expression of total lattice exponent $18/18$, with experimental uncertainties; the charged-mass form carries its scale band. The vertical band is the anchor $\e=|V_{ub}|^{3/10}=0.1869\pm0.0018$. The spread measures the order-one coefficients through the amplification $18/n$ of each estimator.}
\label{fig:estimators}
\end{figure*}
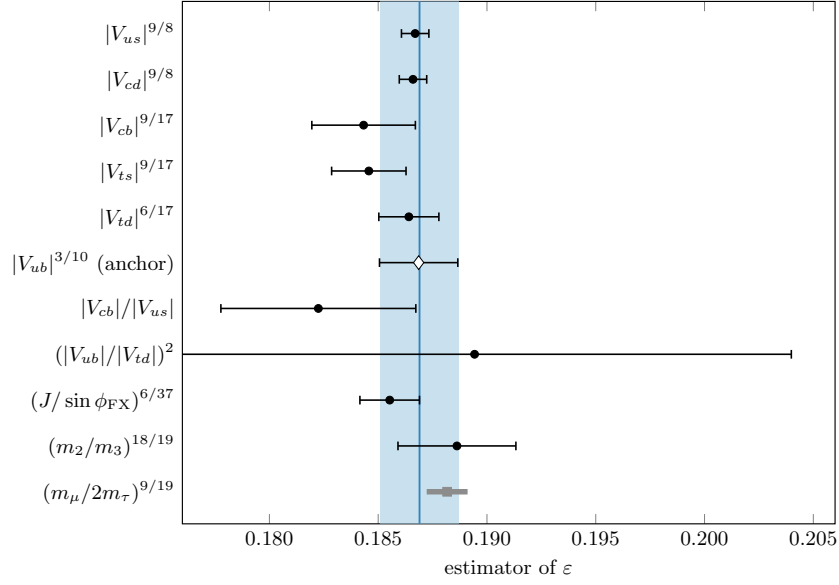

Since a quantity assigned as $X=c\,\e^{n/18}$ returns $X^{18/n}=c^{18/n}\,\e$, each estimator carries its coefficient amplified by $18/n$, a sensitivity factor in the sense that a fractional deviation $\delta$ of the input, whether from its order-one coefficient or from experimental error, enters the estimator as $(18/n)\,\delta$, and the observed spread, $0.1822$ to $0.1894$, a $3.9\%$ band, is the order-one coefficient content of the lattice made visible through the listed amplifications; the ratio family sits low because it inherits the $\sin\theta$ deficit at full strength, while the high-exponent elements are protected. This is also the principled justification of the anchor; among the moduli, $|V_{ub}|^{3/10}$ carries the smallest amplification, $0.300$, so the definition of $\e$ is the one least sensitive to the coefficients it is used to measure. As with Table~\ref{tab:offdiag}, the enumeration re-expresses the existing assignments and claims no additional statistical weight; Fig.~\ref{fig:estimators} displays the convergence.

The same closure permits a change of base. If $X_i=c_i\,\e^{m_i/18}$, then for any base element $B=c_B\,\e^{m_B/18}$ one has $X_i=c_i'\,B^{\,m_i/m_B}$; the integer numerators are invariant, and the denominator becomes the base's own integer. Taking the Cabibbo element as base, $\lambda\equiv\Vus=\e^{16/18}$, the lattice becomes a description in sixteenths,
\begin{equation}
\begin{split}
&\big(|V_{cd}|,\;|V_{cb}|,\;|V_{ts}|,\;|V_{td}|,\;|V_{ub}|\big)\\
&\qquad=\;\lambda^{\,(16,\,34,\,34,\,51,\,60)/16}\\
&\qquad=\;\big(\lambda,\;\lambda^{17/8},\;\lambda^{17/8},\;\lambda^{51/16},\;\lambda^{15/4}\big),
\end{split}
\label{eq:lambdabase}
\end{equation}
with coefficients $(0.9995,\,0.976,\,0.978,\,0.996,\,1.003)$; every off-diagonal modulus is a fractional power of the best-measured element. In this form the lattice makes direct contact with the Wolfenstein expansion \cite{Wolfenstein1983}. Writing $|V_{cb}|=A\lambda^2$, $|V_{td}|=A\lambda^3|1-\bar\rho-i\bar\eta|$, and $|V_{ub}|=A\lambda^3\sqrt{\bar\rho^2+\bar\eta^2}$, the exponents of Eq.~(\ref{eq:lambdabase}) are the statements
\begin{equation}
\begin{gathered}
A=\lambda^{1/8}=0.830\,,\qquad
|1-\bar\rho-i\bar\eta|=\lambda^{1/16}=0.911\,,\\
\sqrt{\bar\rho^2+\bar\eta^2}=\lambda^{5/8}=0.394\,,
\end{gathered}
\label{eq:wolfdict}
\end{equation}
of which two are independent, against the measured $A=0.826\pm0.015$ ($0.3\sigma$), $|1-\bar\rho-i\bar\eta|=0.912\pm0.009$ ($0.1\sigma$), and $\sqrt{\bar\rho^2+\bar\eta^2}=0.387\pm0.010$ ($0.7\sigma$) \cite{PDG2024}. The eighteenths lattice is thus, in base $\lambda$, a fractional-power completion of the Wolfenstein expansion in which the prefactors are not additional parameters but powers of $\lambda$ itself; the residual $0.7\sigma$ in the apex radius is the familiar $\sin\theta$-sector coefficient seen once more. This too is a re-expression, adding no statistical content.

\begin{table*}[t]
\centering
\begin{tabular}{lccc}
\toprule
Quantity & Lattice value ($c_i=1$, $\phifx=\pi/2$) & Data & Coefficient \\
\midrule
$\Vus$ & $0.2219$ & $0.22497\pm0.00068$ & $0.986$ \\
$|V_{cb}|$ & $0.0419$ & $0.0410\pm0.0010$ & $1.023$ \\
$|V_{td}/V_{ts}|$ & $0.2096$ & $0.2085\pm0.0040$ & $1.005$ \\
$J$ & $3.14\times10^{-5}$ & $(3.08\pm0.14)\times10^{-5}$ & $1.018$ \\
$(\alpha,\beta,\gamma)$ & $(88.9^\circ,\,23.0^\circ,\,68.0^\circ)$ & $(84.9^{+5.1}_{-4.5},\,22.2\pm0.7,\,65.4^{+3.8}_{-4.2})^\circ$ & \\
\bottomrule
\end{tabular}
\caption{The CKM matrix obtained from Eq.~(\ref{eq:lattice}) with unit coefficients and $\phifx=\pi/2$, at $\e=|V_{ub}|^{3/10}=0.1869$. $|V_{ub}|$ is the anchor and is excluded from the comparison. The last column lists the order-one coefficient that would bring each quantity to its central value.}
\label{tab:lattice}
\end{table*}
\begin{table*}[t]
\centering
\begin{tabular}{lccccc}
\toprule
Element & Value & Exponent ($\times\tfrac{1}{18}$) & Distance ($\pm\sigma$) & Assignment & Coefficient \\
\midrule
$\Vus$     & $0.22497$ & $16.01$ & $0.01\pm0.03$ & $\e^{16/18}$ & $0.999$ \\
$|V_{cd}|$ & $0.22486$ & $16.01$ & $0.01\pm0.03$ & $\e^{16/18}$ & $0.999$ \\
$|V_{cb}|$ & $0.0410$  & $34.28$ & $0.28\pm0.26$ & $\e^{34/18}$ & $0.975$ \\
$|V_{ts}|$ & $0.04110$ & $34.25$ & $0.25\pm0.19$ & $\e^{34/18}$ & $0.977$ \\
$|V_{td}|$ & $0.00857$ & $51.07$ & $0.07\pm0.23$ & $\e^{51/18}$ & $0.993$ \\
$|V_{ub}|$ & $0.00373$ & $60.00$ & anchor        & $\e^{60/18}$ & $1$ \\
\bottomrule
\end{tabular}
\caption{The six off-diagonal CKM moduli on the eighteenths lattice at $\e=|V_{ub}|^{3/10}=0.1869$. The four exponents are FX arithmetic; $16=17-1$ (interference), $34$ ($2$-$3$ angle), $51=17+34$, $60=26+34$. The table reorganizes the inputs of Table~\ref{tab:inversion} and carries no independent statistical weight; $|V_{cd}|$ follows from $\Vus$ by unitarity, and $|V_{td}|$, $|V_{ts}|$ follow from the FX identities.}
\label{tab:offdiag}
\end{table*}
\begin{table*}[t]
\centering
\begin{tabular}{lccc}
\toprule
Estimator & Value & Exp.\ uncertainty & Amplification $18/n$ \\
\midrule
$\Vus^{9/8}$ & $0.18670$ & $0.00063$ & $1.125$ \\
$|V_{cd}|^{9/8}$ & $0.18660$ & $0.00063$ & $1.125$ \\
$|V_{cb}|^{9/17}$ & $0.18433$ & $0.00238$ & $0.529$ \\
$|V_{ts}|^{9/17}$ & $0.18457$ & $0.00171$ & $0.529$ \\
$|V_{td}|^{6/17}$ & $0.18641$ & $0.00138$ & $0.353$ \\
$|V_{ub}|^{3/10}$ (anchor) & $0.18686$ & $0.00180$ & $0.300$ \\
$|V_{cb}|/\Vus$ & $0.18225$ & $0.00448$ & $1$ \\
$(|V_{ub}|/|V_{td}|)^2$ & $0.18943$ & $0.01456$ & $2$ \\
$(J/\sin\phifx)^{6/37}$ & $0.18553$ & $0.00137$ & $0.162$ \\
$(m_2/m_3)^{18/19}$ & $0.18862$ & $0.00271$ & $0.947$ \\
$(m_\mu/2m_\tau)^{9/19}$ & $0.18817$ & scale band & $0.474$ \\
\bottomrule
\end{tabular}
\caption{Estimators of $\e$; every listed expression has total lattice exponent $18/18$. Uncertainties are experimental only. Each estimator returns $c^{18/n}\e$, so the spread across the table measures the order-one coefficients through the amplification in the last column, the factor $18/n$ by which fractional deviations of the input, in coefficient or in experimental error, are scaled in the estimator. Unitarity partners of the ratio family ($|V_{cb}|/|V_{cd}|$, $|V_{ts}|/\Vus$, $|V_{ts}|/|V_{cd}|$) are omitted; the charged-mass form carries a $\sim0.5\%$ renormalization-scale band in place of an experimental uncertainty.}
\label{tab:estimators}
\end{table*}

One structural property of the lattice deserves emphasis, because it determines where the hypothesis has content and where it does not. The lattice spacing is multiplicative, $\e^{1/18}=0.911$, so neighboring points differ by $9\%$ and every real number lies within $4.6\%$ of some point. For order-one quantities the lattice is dense and the statement ``on-lattice with an order-one coefficient'' is empty. The hypothesis has content only in the hierarchical regime, where exponents are large, integer placement is a genuine constraint, and several independent quantities must land near integers simultaneously. This is the regime of quark mixing; it is not the regime of the large lepton mixing angles, a distinction that shapes Sec.~\ref{sec:pmns}.

The hierarchical regime is not confined to the quark sector, and the lattice admits one clean cross-sector member. The neutrino mass ratio, formed from the measured splittings in the normal ordering with hierarchical $m_1$ \cite{NuFIT2024},
\begin{equation}
\frac{m_2}{m_3}\;=\;\sqrt{\frac{\Delta m_{21}^2}{\Delta m_{31}^2}}\;=\;0.1719\pm0.0026\,,
\end{equation}
carries fitted exponent $18.89$, at distance $0.11$ from the integer, so that
\begin{equation}
\frac{m_2}{m_3}\;=\;\e^{19/18}
\qquad\text{with coefficient }1.010\pm0.018\,,
\label{eq:m23}
\end{equation}
as clean a placement as any of the quark angles. The unrefined integer step, $m_2/m_3=\e^{18/18}$, is rejected at $5.6\sigma$; the half-step is selected here as it is by $\sin\theta_d$. Inverting through the anchor of Eq.~(\ref{eq:epsdef}) gives the direct relation $|V_{ub}|=(m_2/m_3)^{60/19}$, satisfied at the $3\%$ level; a CKM element and the solar-to-atmospheric splitting ratio are tied with no lepton mixing angle involved, which is where the regime argument permits a cross-sector relation to live. The same ratio extends to the reactor angle through a fixed dressing, $\sin\theta_{13}=\tfrac{\sqrt{3}}{2}\,m_2/m_3=\tfrac{\sqrt{3}}{2}\,\e^{19/18}$ with coefficient $1.006$, a relation developed in Sec.~\ref{sec:pmns}. The ratio $m_2/m_3$ is the right object for these tests on renormalization grounds as well; for a hierarchical normal-ordered spectrum the light neutrino masses run essentially multiplicatively below the seesaw scale, so the ratio is scale-stable at the sub-percent level.

The hierarchical limit is not innocuous at this precision; under the geometric completion of Sec.~\ref{sec:pmns}, $m_1=m_2^2/m_3$, the ratio shifts by $+1.5\%$ to $0.1745$, moving the fitted exponent to $18.74$ and the coefficient to $1.025\pm0.018$, within $1.4\sigma$ of the assignment either way; we quote the hierarchical limit throughout, and the shift bounds the $m_1$ sensitivity of the placement. The relation is also a normal-ordering statement; in the inverted ordering the two measurable masses are nearly degenerate and offer no small ratio to tie to $\e$, so the framework selects normal ordering, and the mass-ordering determination is itself among its tests.

The charged-lepton mass ratios provide the cautionary counterpart. Because these ratios run, a lattice claim about them requires a stated scale; we adopt $M_Z$, the standard reference point for flavor comparisons, with $m_e(M_Z)=0.48657$~MeV, $m_\mu(M_Z)=102.718$~MeV, and $m_\tau(M_Z)=1746.24$~MeV \cite{XZZ2008}. Table~\ref{tab:inventory} collects the full lattice inventory. At $M_Z$ the charged-lepton ratios sit at distances $0.40$ and $0.44$ from the nearest integers, essentially mid-lattice; they are off-lattice and are excluded. Their pole-mass distances, $0.29$ and $0.21$, quantify the scale sensitivity, about $0.1$ lattice unit per percent of running, and illustrate why mass-ratio placements without a stated scale carry no weight. The geometric-mean combination $\sqrt{m_e m_\tau}/m_\mu$, markedly less scale-sensitive than the individual ratios, provides the robust summary; its fitted exponent sits at distance $0.48$ from the nearest integer at $M_Z$ and $0.46$ at the pole masses, essentially the mid-lattice maximum at either scale. The inventory thus contains rejections as well as placements, which is what gives the criterion its content; only the scale-stable $m_2/m_3$ joins the quark angles, and it does so at their level of quality.

\begin{table*}[t]
\centering
\begin{tabular}{lcccccc}
\toprule
Quantity & Value & Exponent ($\times\tfrac{1}{18}$) & Distance & Assignment & Coefficient & Status \\
\midrule
$\sin\theta_u$ & $0.0906$ & $25.77$ & $0.23$ & $\e^{26/18}$ & $1.022$ & on-lattice \\
$\sin\theta_d$ & $0.2042$ & $17.05$ & $0.05$ & $\e^{17/18}$ & $0.995$ & on-lattice \\
$\sin\theta$   & $0.0412$ & $34.23$ & $0.23$ & $\e^{34/18}$ & $0.979$ & on-lattice \\
$m_2/m_3$      & $0.1719\pm0.0026$ & $18.89$ & $0.11$ & $\e^{19/18}$ & $1.010\pm0.018$ & on-lattice \\
$m_\mu/m_\tau$ ($M_Z$) & $0.05882$ & $30.40$ & $0.40$ & none & & off-lattice \\
$m_e/m_\mu$ ($M_Z$)    & $0.004737$ & $57.44$ & $0.44$ & none & & off-lattice \\
\bottomrule
\end{tabular}
\caption{Lattice inventory. Fitted exponents are $18\ln x/\ln\e$ with $\e=|V_{ub}|^{3/10}=0.1869$; the distance is to the nearest integer, with $0.5$ the maximum possible. $|V_{ub}|$ itself is the anchor, with exponent $60$ by construction, and is not a test. Charged-lepton ratios are evaluated at $M_Z$ \cite{XZZ2008}; their pole-mass distances, $0.29$ and $0.21$, quantify the scale sensitivity discussed in the text.}
\label{tab:inventory}
\end{table*}

\section{Quark mass ratios at \texorpdfstring{$M_Z$}{MZ} and a mass-ratio construction of the CKM matrix}
\label{sec:qmass}

The lattice extends to the quark mass spectrum, and the extension closes the circle the Cabibbo-angle literature opened; it produces an expression for the full CKM matrix in terms of quark mass ratios alone. All ratios are evaluated at $M_Z$, using $m_d/m_s=0.0503\pm0.0007$ (scale-invariant among the light quarks) and $m_s(M_Z)=52.9$~MeV, $m_c(M_Z)=0.620$~GeV, $m_b(M_Z)=2.86$~GeV, $m_t(M_Z)=168.3$~GeV \cite{XZZ2008}, consistent with the PDG inputs evolved to $M_Z$ \cite{PDG2024}. Ratios of same-charge quarks at a common scale are protected by the flavor-universal QCD anomalous dimension, so these placements, unlike the charged-lepton ones above, are stable under the choice of reference scale up to threshold matching and electroweak corrections. Table~\ref{tab:qmass} places the six independent ratios on the lattice.

\begin{table*}[t]
\centering
\begin{tabular}{lccccc}
\toprule
Ratio & Value at $M_Z$ & Exponent & Distance ($\pm\sigma$) & Assignment & Coefficient \\
\midrule
$m_d/m_s$ & $0.0503$ & $32.09$ & $0.09\pm0.15$ & $\e^{32/18}$ & $0.992$ \\
$m_s/m_b$ & $0.0185$ & $42.82$ & $0.18\pm0.17$ & $\e^{43/18}$ & $1.017$ \\
$m_c/m_t$ & $0.00368$ & $60.13$ & $0.13\pm0.32$ & $\e^{60/18}$ & $0.988$ \\
$m_u/m_c$ & $0.00197$ & $66.84$ & $0.16\pm0.54$ & ($\e^{67/18}$) & $1.015$ \\
$m_c/m_b$ & $0.2168$ & $16.41$ & $0.41\pm0.32$ & ($\e^{16/18}$) & $0.963$ \\
$m_u/m_d$ & $0.460$ & $8.34$ & $0.34\pm0.43$ & unconstraining & \\
\bottomrule
\end{tabular}
\caption{Quark mass ratios at $M_Z$ on the eighteenths lattice. The first three carry the construction of Eq.~(\ref{eq:massconstruction}); $m_u/m_c$ is weakly constrained and its integer plays no role below; $m_c/m_b$ is consistent with the integer $16$ only within its sizable uncertainty and likewise plays no role below; $m_u/m_d$ is too poorly known to test.}
\label{tab:qmass}
\end{table*}

The three well-placed ratios carry integers $(32,\,43,\,60)$, and the four FX integers are their combinations,
\begin{equation}
\begin{gathered}
16=\tfrac{32}{2},\qquad 17=60-43,\\
26=2\times43-60,\qquad 34=2\,(60-43),
\end{gathered}
\end{equation}
with the consistency check $m_d/m_b=\e^{75/18}$, $75=32+43$, satisfied at distance $0.09$. Because the combinations have unit coefficients, they are equivalent to three parameter-free relations among measured quantities, independent of $\e$ altogether:
\begin{equation}
\begin{gathered}
\frac{m_d}{m_s}=\lambda^2\,,\qquad
\frac{m_s}{m_b}=\sin\theta_d\,\sin\theta_u\,,\\
\frac{m_c}{m_t}=\sin\theta_u\,\sin\theta=|V_{ub}|\,,
\end{gathered}
\label{eq:massrelations}
\end{equation}
satisfied at $0.7\%$, $0.03\%$, and $1.2\%$ respectively; against the $\sim1.6\%$ uncertainty of the mass side, the second relation is exact to $0.02\sigma$. The first is the Gatto--Sartori--Tonin relation \cite{GST1968} in its down-sector form, and the lattice locates it at the observable level; $m_d/m_s$ carries exponent $32=2\times16$, twice the interference-shifted Cabibbo exponent, not $2\times17$, so the mass ratio pairs with $\lambda$ rather than with $\tan\theta_d$, and the naive texture identification $\tan\theta_d=\sqrt{m_d/m_s}$ is disfavored at $3\sigma$ by the same data.

\begin{figure*}[t]
\centering
\resizebox{0.86\textwidth}{!}{%
\begin{tikzpicture}[x=1.32cm,y=0.85cm,
  bx/.style={draw,fill=white,inner xsep=5pt,inner ysep=4pt,font=\small},
  tg/.style={fill=white,inner sep=1.5pt,font=\scriptsize},
  ar/.style={-{Latex[length=2.4mm]},black!70,line width=0.7pt}]
\node[bx] (t1) at (1.5,5.7) {$m_d/m_s\ \ [32]$};
\node[bx] (t2) at (4.6,5.7) {$m_s/m_b\ \ [43]$};
\node[bx] (t3) at (7.7,5.7) {$m_c/m_t\ \ [60]$};
\node[bx] (m0) at (1.3,3.4) {$\lambda=|V_{us}|\ \ [16]$};
\node[bx] (m1) at (3.7,3.4) {$\sin\theta_d\ \ [17]$};
\node[bx] (m2) at (5.7,3.4) {$\sin\theta_u\ \ [26]$};
\node[bx] (m3) at (7.7,3.4) {$\sin\theta\ \ [34]$};
\node[bx,fill=cboxfill] (mp) at (9.55,3.4) {$\phi_{\rm FX}=\pi/2$};
\node[bx] (b1) at (1.5,1.1) {$|V_{us}|\ \ [16]$};
\node[bx] (b2) at (4.2,1.1) {$|V_{cb}|\ \ [34]$};
\node[bx] (b3) at (6.4,1.1) {$|V_{td}|\ \ [51]$};
\node[bx] (b4) at (8.7,1.1) {$|V_{ub}|\ \ [60]$};
\draw[ar] (t1) -- (m0);
\foreach \s in {t2,t3}{ \draw[ar] (\s) -- (m1); \draw[ar] (\s) -- (m2); \draw[ar] (\s) -- (m3); }
\draw[ar] (m0) -- (b1);
\draw[ar] (m3) -- (b2); \draw[ar] (m1) -- (b3); \draw[ar] (m3) -- (b3);
\draw[ar] (m2) -- (b4); \draw[ar] (m3) -- (b4);
\begin{scope}[on background layer]
\draw[-{Latex[length=2.4mm]},cblue,line width=0.7pt] (mp.south) to[bend left=14] (b1.north east);
\end{scope}
\node[tg] at (1.3,4.62) {$16=32/2$};
\node[tg] at (3.7,4.62) {$17=60-43$};
\node[tg] at (5.55,4.62) {$26=2\cdot43-60$};
\node[tg] at (7.9,4.62) {$34=2(60-43)$};
\node[tg] at (4.2,2.32) {$34$};
\node[tg] at (6.4,2.32) {$51=17+34$};
\node[tg] at (8.7,2.32) {$60=26+34$};
\node[tg,text=cblue] at (5.6,0.30) {interference at $\phi_{\rm FX}=\pi/2$: Cabibbo exponent $17\to16$};
\end{tikzpicture}}
\caption{The mass-ratio construction of the CKM matrix as exponent arithmetic; brackets give the lattice integer of each quantity. The three on-lattice quark mass ratios (top) generate the Cabibbo base $\lambda$ and the three FX angles (middle) through the combinations of the edge labels, and the FX sums generate the off-diagonal moduli (bottom); $|V_{cd}|$ and $|V_{ts}|$, not drawn, carry the same integers, $16$ and $34$, as their unitarity partners. The single angular input $\phifx=\pi/2$ supplies the interference that shifts the Cabibbo exponent from $17$ to $16$ and, together with the angles, the CP-violating invariants.}
\label{fig:construction}
\end{figure*}
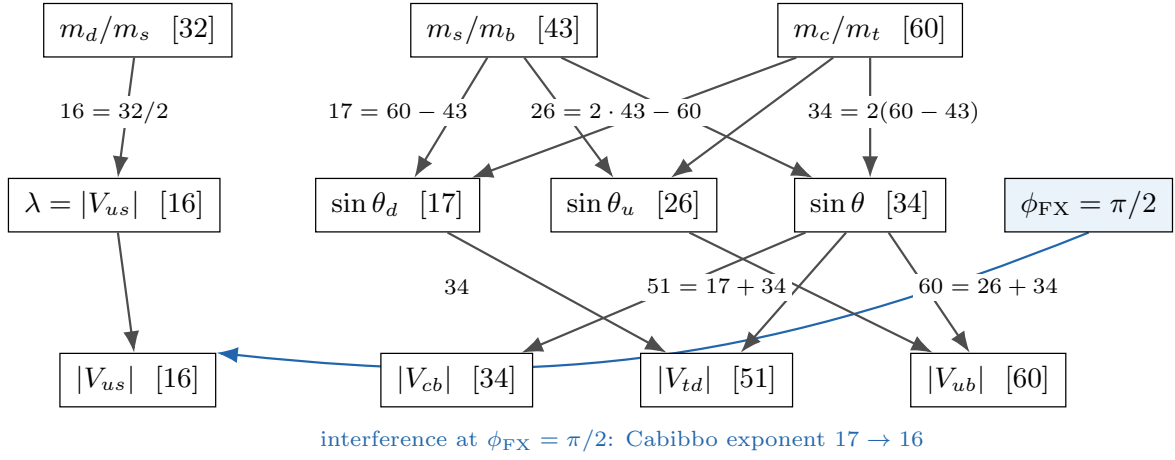

Inverting Eq.~(\ref{eq:massrelations}) gives the construction. With $R_1\equiv m_s/m_b$ and $R_2\equiv m_c/m_t$,
\begin{equation}
\begin{gathered}
\lambda=\sqrt{\frac{m_d}{m_s}}\,,\qquad
\sin\theta_d=\frac{R_2}{R_1}\,,\\
\sin\theta_u=\frac{R_1^2}{R_2}\,,\qquad
\sin\theta=\left(\frac{R_2}{R_1}\right)^{\!2},\\
\phifx=\frac{\pi}{2}\,,
\end{gathered}
\label{eq:massconstruction}
\end{equation}
and Eq.~(\ref{eq:fxmatrix}) then delivers the full matrix from three mass ratios and the maximal phase. Numerically the angles come out at $(-2.4\%,\,+2.5\%,\,-3.7\%)$ from their measured values, and the derived observables at $|V_{cb}|=0.0395$, $|V_{td}/V_{ts}|=0.2033$, $J=2.8\times10^{-5}$, and $(\alpha,\beta,\gamma)=(88.9^\circ,24.7^\circ,66.4^\circ)$, each within the compounded order-one coefficients; $|V_{ub}|=m_c/m_t$ holds identically in this algebra and at $1.2\%$ in data. Figure~\ref{fig:construction} draws the construction as exponent arithmetic.

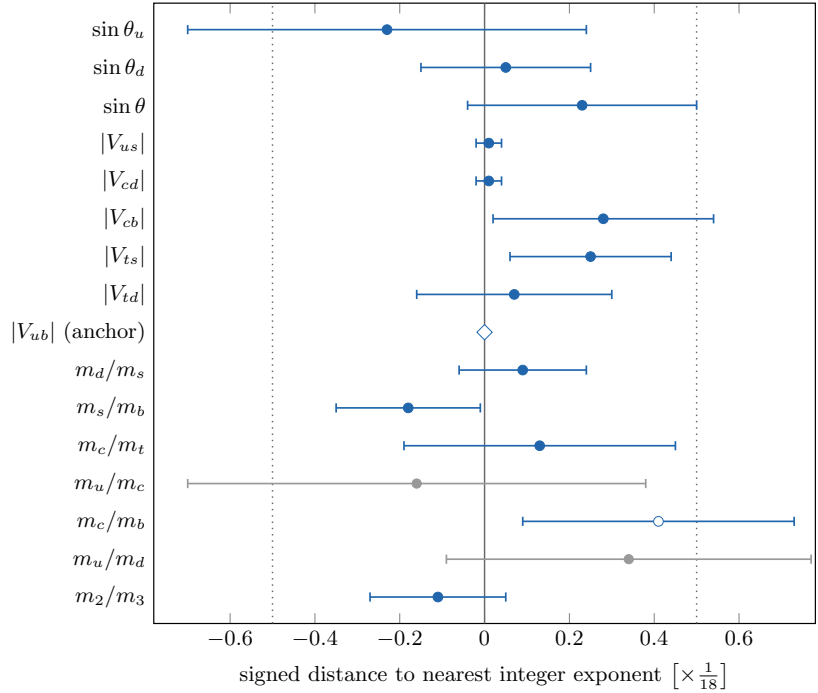
\begin{figure*}[t]
\centering
\resizebox{0.60\textwidth}{!}{%
\begin{tikzpicture}
\begin{axis}[width=11.2cm,height=10.6cm,
  xmin=-0.78,xmax=0.78,ymin=-0.7,ymax=15.7,
  xlabel={signed distance to nearest integer exponent  $\left[\times\frac{1}{18}\right]$},
  ytick={15,14,13,12,11,10,9,8,7,6,5,4,3,2,1,0},yticklabels={{$\sin\theta_u$},{$\sin\theta_d$},{$\sin\theta$},{$|V_{us}|$},{$|V_{cd}|$},{$|V_{cb}|$},{$|V_{ts}|$},{$|V_{td}|$},{$|V_{ub}|$ (anchor)},{$m_d/m_s$},{$m_s/m_b$},{$m_c/m_t$},{$m_u/m_c$},{$m_c/m_b$},{$m_u/m_d$},{$m_2/m_3$}},
  ytick style={draw=none}, tick label style={font=\small},
  axis line style={line width=0.5pt}]
\draw[black!60,line width=0.6pt] (axis cs:0,-0.7) -- (axis cs:0,15.7);
\draw[black!60,line width=0.6pt,dotted] (axis cs:-0.5,-0.7) -- (axis cs:-0.5,15.7);
\draw[black!60,line width=0.6pt,dotted] (axis cs:0.5,-0.7) -- (axis cs:0.5,15.7);
\addplot[only marks,mark=*,mark size=2pt,color=cblue,
  error bars/.cd,x dir=both,x explicit,
  error bar style={line width=0.7pt},error mark options={line width=0.7pt,mark size=1.9pt,rotate=90}]
  coordinates { (-0.23,15) +- (0.47,0) (+0.05,14) +- (0.20,0) (+0.23,13) +- (0.27,0) (+0.01,12) +- (0.03,0) (+0.01,11) +- (0.03,0) (+0.28,10) +- (0.26,0) (+0.25,9) +- (0.19,0) (+0.07,8) +- (0.23,0) (+0.09,6) +- (0.15,0) (-0.18,5) +- (0.17,0) (+0.13,4) +- (0.32,0) (-0.11,0) +- (0.16,0) };
\addplot[only marks,mark=*,mark size=2pt,mark options={fill=white,draw=cblue},color=cblue,
  error bars/.cd,x dir=both,x explicit,
  error bar style={line width=0.7pt},error mark options={line width=0.7pt,mark size=1.9pt,rotate=90}]
  coordinates { (+0.41,2) +- (0.32,0) };
\addplot[only marks,mark=*,mark size=1.9pt,color=black!40,
  error bars/.cd,x dir=both,x explicit,
  error bar style={line width=0.7pt},error mark options={line width=0.7pt,mark size=1.9pt,rotate=90}]
  coordinates { (-0.16,3) +- (0.54,0) (+0.34,1) +- (0.43,0) };
\node[diamond,draw=cblue,fill=white,minimum size=6.5pt,inner sep=0pt] at (axis cs:0.0,7) {};
\end{axis}
\end{tikzpicture}}
\caption{Lattice placements: signed distance of the fitted exponent to the nearest integer, in units of $1/18$, with its uncertainty; uncertainties are propagated from the experimental inputs alone, without the anchor correlation. Filled blue circles are the lattice placements; the open diamond is the anchor $|V_{ub}|$, at zero by construction; the open circle is the marginal $m_c/m_b$ (see text); gray entries are too poorly determined to test. Dotted lines at $\pm0.5$ mark the maximal possible distance. The charged-lepton entries are tabulated in Table~\ref{tab:inventory}.}
\label{fig:lattice}
\end{figure*}

The exclusions are as informative as the placements. The up-sector ratio $m_u/m_c$ appears nowhere in Eq.~(\ref{eq:massconstruction}); the classic identification $\sin\theta_u=\sqrt{m_u/m_c}$ fails by the well-known factor of two ($0.044$ against $0.091$), and the lattice resolves the failure by exclusion rather than repair, replacing it with $\sin\theta_u=R_1^2/R_2$ at $2.5\%$. The ratio $m_c/m_b$, at distance $0.41\pm0.32$, is consistent with the integer $16$ only within its sizable uncertainty and enters the construction nowhere; $m_u/m_d$ is too poorly known to participate. The lattice thus selects which mass ratios enter the mixing and which do not, a discrimination no texture ansatz supplies. Figure~\ref{fig:lattice} collects the placements in a single view, with a uniform classification: filled circles mark entries whose integers are fixed by the closed arithmetic of the construction and whose central distances lie at or below $0.3$; gray marks ratios without discriminating power (uncertainty at or above $0.4$) that play no role in the construction; and the open circle marks the marginal $m_c/m_b$, central distance $0.41$ with uncertainty $0.32$, consistent with $16$ at $1.3\sigma$ yet outside the placement band, entering the construction and the statistics of Sec.~\ref{sec:stats} nowhere.

\section{Statistical weight}
\label{sec:stats}

A pattern paper owes the reader an accounting of chance. We give it here for each structure separately, stating the look-elsewhere effects rather than absorbing them.

For the eighteenths lattice, the relevant statistic after the anchor of Eq.~(\ref{eq:epsdef}) is the placement of the three fitted exponents of Eq.~(\ref{eq:fitexp}). Under the null hypothesis of no lattice, each distance to the nearest integer is uniform on $[0,0.5]$; the probability that all three land within the largest observed distance, $0.23$, is $(0.46)^3\simeq0.10$. The quark mass ratios of Sec.~\ref{sec:qmass} add three further independent placements, at distances $0.09$, $0.18$, and $0.13$, with joint chance probability $(0.36)^3\simeq0.05$; and the unit-coefficient product relations they imply, Eq.~(\ref{eq:massrelations}), carry no adjustable constants at all, with $m_s/m_b=\sin\theta_d\sin\theta_u$ holding at $0.02\sigma$, the sharpest retrodictive agreement in this paper. The choice of denominator is itself a look-elsewhere effect, but a bounded one; as shown in Sec.~\ref{sec:ninths}, eighteen is the minimal denominator consistent with the data, since the coarser ninths lattice is rejected at the $5\sigma$ level by $\sin\theta_d$ alone, and no finer denominator is invoked anywhere in the analysis. A skeptic granting trials over denominators would still dilute the probabilities toward the tens of percent. We therefore characterize the multiplicative lattice, on its own, as suggestive and not significant; its scientific weight lies in the compact structure it exposes, the closed form of Eq.~(\ref{eq:J}), and the relations of Sec.~\ref{sec:map}, not in a small $p$-value.

For the $\pi/24$ triangle of Sec.~\ref{sec:triangle}, closure fixes one angle, so the pattern has two independent matches. With the observed deviations, $0.3^\circ$ in $\beta$ and $2.1^\circ$ in $\gamma$, the chance probability for two uniform angles is $(0.3/3.75)(2.1/3.75)\simeq0.04$. This is again modest, and we present it as such.

The conclusion we draw from this accounting is methodological; retrodictive agreement, at these probabilities, cannot carry the case, and the case therefore rests on the prospective tests. Those are sharp. The lattice-plus-maximal-phase construction places $\gamma$ at $67.5^\circ$--$68.0^\circ$, between one and two degrees above the current central values, and the LHCb and Belle~II programs are driving $\gamma$ toward degree-level precision; the same construction fixes $\beta$ within a degree of its present central value; and the leptonic extension of Sec.~\ref{sec:pmns} predicts $\delta_{\rm CP}$ outright. The patterns will be confirmed or excluded by measurement within the decade, which is the standard an empirical proposal should be held to.

\section{The \texorpdfstring{$\pi/24$}{pi/24} unitarity triangle}
\label{sec:triangle}

Convention-independent statements are carried by the unitarity triangle $V_{ud}V_{ub}^*+V_{cd}V_{cb}^*+V_{td}V_{tb}^*=0$, whose angles satisfy $\alpha+\beta+\gamma=180^\circ$ identically. On the $\pi/24$ lattice the data-compatible assignment is
\begin{equation}
(\alpha,\beta,\gamma)\;=\;(12,3,9)\times7.5^\circ\;=\;(90^\circ,\,22.5^\circ,\,67.5^\circ),
\label{eq:halftick}
\end{equation}
with closure automatic. (A coarser unit of $15^\circ$ would force $\beta=30^\circ$, excluded by the measured $\beta=22.2^\circ\pm0.7^\circ$; the half-unit is the coarsest commensurate choice the data allow.) Against current data the deviations of Eq.~(\ref{eq:halftick}) are $0.4\sigma$ in $\beta$, $0.6\sigma$ in $\gamma$ relative to $65.4^{+3.8}_{-4.2}$ \cite{LHCbGamma2021}, and $1.0\sigma$ in $\alpha$; the pattern is consistent everywhere, with the chance probability quantified in Sec.~\ref{sec:stats} and the decisive test supplied by the coming precision on $\gamma$.

Table~\ref{tab:quantization} collects the angular quantization in one view; every angle of the analysis carries an integer $n$ on the $7.5^\circ$ lattice, measured on the quark side and predictive on the lepton side. The Cabibbo angle is listed as the instructive exclusion; $\theta_C=13.00^\circ$ lies far from $2\times7.5^\circ$, and the angle is carried instead by the multiplicative structure at the eighteenths integer $n=16$, $\theta_C\simeq\e^{16/18}=\sqrt[18]{\e^{16}}$, Eq.~(\ref{eq:outputs}), the division of labor of Sec.~\ref{sec:discussion} in miniature. A companion paper employs the coarser $\pi/12$ quantization for trimaximal first-column lepton mixing \cite{Pi12Model}. Read this way, the angular structure is a digital clock;\footnote{The name describes a quantization of measured angles and is unrelated to the clockwork mechanism \cite{Clockwork2016,ClockworkFlavor2018}, which generates exponential hierarchies from chains of order-one couplings; contact with that framework, if any, would enter only at the dynamical level from which this analysis abstains.} every angle of the analysis sits on one of the $48$ ticks of a $7.5^\circ$ dial, and the octant question is which of the two ticks, $14$ or $7$, the leptonic apex occupies.

\begin{table*}[t]
\centering
\begin{tabular}{lccll}
\toprule
Angle & $n$ & $n\times7.5^\circ$ & Value & Status \\
\midrule
$\alpha$ & $12$ & $90^\circ$ & $(84.9^{+5.1}_{-4.5})^\circ$ & consistent, $1.0\sigma$ \\
$\beta$ & $3$ & $22.5^\circ$ & $22.2^\circ\pm0.7^\circ$ & consistent, $0.4\sigma$ \\
$\gamma$ & $9$ & $67.5^\circ$ & $(65.4^{+3.8}_{-4.2})^\circ$ & consistent, $0.6\sigma$; decisive test \\
$\phifx$ & $12$ & $90^\circ$ & $92.3^\circ\pm2.7^\circ$ & maximal phase, $0.9\sigma$ \\
$|\alpha_\ell|$ & $14$ & $105^\circ$ & & prediction (upper octant) \\
$|\beta_\ell|$ & $3$ & $22.5^\circ$ & & prediction; shared with $\beta$ \\
$|\gamma_\ell|$ & $7$ & $52.5^\circ$ & & prediction (upper octant) \\
$\theta_C$ & $(2)$ & $(15^\circ)$ & $13.00^\circ\pm0.04^\circ$ & excluded; multiplicative, $\theta_C\simeq\sqrt[18]{\e^{16}}$ \\
\bottomrule
\end{tabular}
\caption{The angular quantization: angles as integer multiples of $\pi/24=7.5^\circ$. Quark-sector entries are measured; the leptonic entries are the predictions of Eq.~(\ref{eq:pmnspred}), with the octant exchange interchanging $n=14$ and $n=7$. The $\phifx$ row extends the quantization from rephasing invariants to the FX phase through the map $\alpha\simeq\phifx$. The Cabibbo angle is the instructive exclusion; it sits far off the angular lattice and belongs to the multiplicative structure, Eq.~(\ref{eq:outputs}).}
\label{tab:quantization}
\end{table*}

\section{The map between the lattices}
\label{sec:map}

The triangle generated by the eighteenths lattice in Table~\ref{tab:lattice}, $(88.9^\circ,\,23.0^\circ,\,68.0^\circ)$, lies within about a degree of the $\pi/24$ assignment of Eq.~(\ref{eq:halftick}), and the correspondence follows from two short relations.

First, the FX phase maps onto the triangle angle $\alpha$ up to a small correction. At the pure-lattice point $\alpha=88.9^\circ$ against $\phifx=90^\circ$, and at the measured parameters $\alpha=91.2^\circ$ against $\phifx=92.3^\circ$, a correction near $-1.1^\circ$ in both cases. The lattice statement $\alpha=12\times7.5^\circ$ is therefore the image of the maximal phase.

Second, the exponent difference in Eq.~(\ref{eq:lattice}) is $26/18-17/18=1/2$, so at maximal phase
\begin{equation}
\tan\beta \;\simeq\; \frac{s_u}{s_d} \;=\; \sqrt{\e}\,,
\qquad \gamma \;\simeq\; 90^\circ-\beta\,,
\label{eq:tanbeta}
\end{equation}
up to the degree-level corrections visible in Table~\ref{tab:lattice}. The half-integer power arises from the eighteenths lattice, not from any angular input. Equation~(\ref{eq:tanbeta}) makes the relation between the two structures exact; the $\pi/24$ triangle is recovered when $\sqrt{\e}=\tan22.5^\circ=\sqrt{2}-1$, that is, at
\begin{equation}
\e^{*} \;=\; 3-2\sqrt{2} \;=\; 0.17157\,.
\label{eq:epsstar}
\end{equation}
This value is excluded by the moduli; at $\e=\e^{*}$ with unit coefficients the same construction gives $|V_{cb}|=0.0357$ and $\Vus=0.204$, low by $13\%$ and $9\%$ respectively. The eighteenths lattice at $\e=0.187$ instead gives $\beta=23.0^\circ$ ($\tan^{-1}\sqrt{\e}=23.4^\circ$ at leading order), one degree above the half-unit value and $1.2\sigma$ above the measured $22.2^\circ\pm0.7^\circ$, a residue absorbed by an order-one coefficient of $0.94$ on $s_u/s_d$.

\begin{figure*}[t]
\centering
\resizebox{0.72\textwidth}{!}{%
\begin{tikzpicture}
\begin{axis}[name=main,width=12.4cm,height=8.8cm,
  xmin=-0.08,xmax=1.06,ymin=-0.06,ymax=0.52,
  xlabel={$\bar\rho$},ylabel={$\bar\eta$},
  tick label style={font=\small},axis line style={line width=0.5pt},
  legend style={at={(0.30,0.10)},anchor=south west,draw=none,fill=none,
    font=\normalsize,legend cell align=left}]
\draw[black!40,line width=0.5pt] (axis cs:-0.08,0) -- (axis cs:1.06,0);
\addplot[color=cblue,line width=1.1pt] coordinates {(0,0) (0.14645,0.35355) (1,0) (0,0)};
\addlegendentry{$\pi/24$: $(90^\circ,22.5^\circ,67.5^\circ)$}
\addplot[color=cgreen,dashed,line width=1.1pt] coordinates {(0,0) (0.14639,0.36233) (1,0) (0,0)};
\addlegendentry{$\varepsilon$ lattice: $(88.9^\circ,23.0^\circ,68.0^\circ)$}
\addplot[color=black!35,line width=0.5pt,forget plot] coordinates {(0,0) (0.15743,0.34385)};
\addplot[color=black!35,line width=0.5pt,forget plot] coordinates {(1,0) (0.15743,0.34385)};
\addplot[only marks,mark=*,mark size=1.8pt,color=black,
  error bars/.cd,x dir=both,x explicit,y dir=both,y explicit,
  error bar style={line width=0.7pt},error mark options={line width=0.7pt,mark size=1.9pt,rotate=90}]
  coordinates { (0.15743,0.34385) += (0.03103,0.02016) -= (0.02727,0.02006) };
\addlegendentry{measured apex}
\coordinate (zUL) at (axis cs:0.115,0.385);
\coordinate (zLL) at (axis cs:0.115,0.315);
\draw[black!45,line width=0.5pt] (axis cs:0.115,0.315) rectangle (axis cs:0.20,0.385);
\end{axis}
\begin{axis}[name=ins,at={(main.north east)},anchor=north east,xshift=-6pt,yshift=-6pt,
  width=4.9cm,height=4.1cm,xmin=0.115,xmax=0.20,ymin=0.315,ymax=0.385,
  tick label style={font=\tiny},axis line style={line width=0.4pt},
  xtick={0.12,0.14,0.16,0.18,0.20},ytick={0.32,0.34,0.36,0.38}]
\addplot[only marks,mark=triangle*,mark size=2.6pt,color=cblue] coordinates {(0.14645,0.35355)};
\addplot[only marks,mark=triangle*,mark size=2.6pt,color=cgreen] coordinates {(0.14639,0.36233)};
\addplot[only marks,mark=*,mark size=1.8pt,color=black,
  error bars/.cd,x dir=both,x explicit,y dir=both,y explicit,
  error bar style={line width=0.7pt},error mark options={line width=0.7pt,mark size=1.9pt,rotate=90}]
  coordinates { (0.15743,0.34385) += (0.03103,0.02016) -= (0.02727,0.02006) };
\end{axis}
\draw[black!45,line width=0.5pt] (zUL) -- (ins.north west);
\draw[black!45,line width=0.5pt] (zLL) -- (ins.south west);
\end{tikzpicture}}
\caption{The CKM unitarity triangle in the $(\bar\rho,\bar\eta)$ plane: the measured apex with its uncertainties, the $\pi/24$ assignment $(90^\circ,22.5^\circ,67.5^\circ)$, drawn from its angles with apex at $(\bar\rho,\bar\eta)=(0.146,\,0.354)$, and the triangle generated by the $\e$ lattice at maximal phase. The three apexes lie within $0.02$ of one another and overlap at full scale; the inset magnifies the apex region.}
\label{fig:triangle}
\end{figure*}
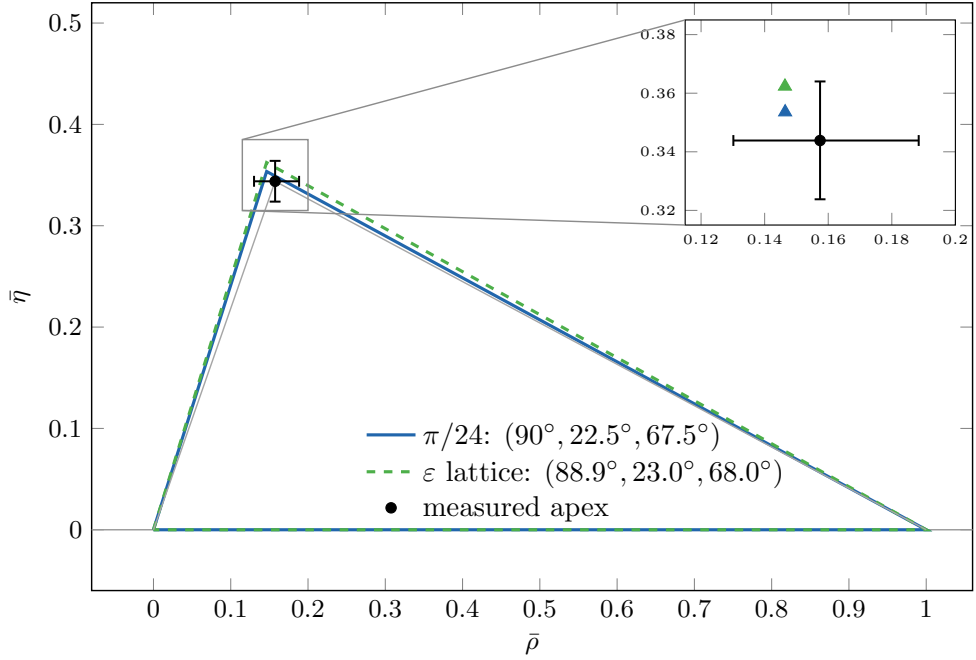

The interpretation we adopt is that the $\pi/24$ triangle is not evidence for an angular symmetry; it is the angular approximation, good to about a degree, of the multiplicative lattice evaluated at maximal phase. The two readings are indistinguishable at current precision and separate in principle; the angular lattice predicts $\beta\to22.5^\circ$ exactly, while the eighteenths lattice predicts $\tan\beta=c\,\sqrt{\e}$ with $c$ of order unity and $\e$ fixed independently by the moduli. Degree-level precision on $\beta$ and $\gamma$ cannot by itself resolve an order-one coefficient, so the sharper discrimination runs through the moduli, which already prefer $\e=0.187$ over $\e^{*}$ decisively. In this reading the Dirac phase and the Cabibbo angle are outputs of the multiplicative structure,
\begin{equation}
\begin{gathered}
\delta \;\simeq\; \gamma \;\simeq\; 90^\circ - \tan^{-1}\!\big(c\,\sqrt{\e}\,\big)
\;\approx\;65^\circ\text{--}68^\circ,\\
\theta_C \;\simeq\; \e^{8/9},
\end{gathered}
\label{eq:outputs}
\end{equation}
and the additive $\pi/24$ pattern is their commensurate summary. Figure~\ref{fig:triangle} displays the three apexes together.

\section{A \texorpdfstring{$23^\circ$}{23-Degree} Theorem for \texorpdfstring{$\beta$}{Beta}}
\label{sec:betatheorem}

The content of the map admits a compact statement. Postulate: the three FX angles are unit-coefficient eighteenth powers of a single parameter, $\sin\theta_u=\e^{26/18}$, $\sin\theta_d=\e^{17/18}$, $\sin\theta=\e^{34/18}$, the phase is maximal, $\phifx=\pi/2$, and $\e$ is fixed by the moduli; no angular information enters. Theorem: the unitarity-triangle angle $\beta$ is then determined, $\tan\beta=\sqrt{\e}$ at leading order, and $\beta=23.0^\circ$ from the full matrix at $\e=0.187$.

The proof is the chain already established. First, $\beta$ is the rephasing invariant $\arg\!\big({-V_{cd}V_{cb}^*/V_{td}V_{tb}^*}\big)$, so the postulate suffices to compute it. Second, in the FX form of Eq.~(\ref{eq:fx}) at maximal phase the invariant reduces to $\tan\beta\simeq s_u/s_d$, Eq.~(\ref{eq:tanbeta}), up to the degree-level corrections of Table~\ref{tab:lattice}. Third, the postulate sets $s_u/s_d=\e^{(26-17)/18}=\e^{1/2}$, the half-integer power arising as an exponent difference, so $\tan\beta=\sqrt{\e}=0.432$ and $\beta=23.4^\circ$ at leading order, sharpened to $23.0^\circ$ by the full matrix. The invariant content of the theorem is the form $\tan\beta=\sqrt{\e}$, which follows from the exponent difference alone and survives order-one coefficients as $\tan\beta=c\,\sqrt{\e}$; the number $23.0^\circ$ is its evaluation at unit coefficients.

Three corollaries follow. The theorem lands within half a degree of the clock tick $3\times7.5^\circ=22.5^\circ$, with equality exactly at $\e^{*}=3-2\sqrt{2}$, Eq.~(\ref{eq:epsstar}), a value the moduli exclude; the clock's $\beta$ is thus the angular approximation of a multiplicative theorem. Against data, the measured $\beta=22.2^\circ\pm0.7^\circ$ sits $1.2\sigma$ below, the residue an order-one coefficient of $0.94$ on $s_u/s_d$. And through the shared $\beta=\beta_\ell$ of Sec.~\ref{sec:pmns}, the theorem's angle is the one number the quark and lepton sectors hold in common, so the sharpest single confrontation of the multiplicative structure with the lepton sector runs through this angle.

A further corollary organizes the remaining angles into small integer ratios. In the quark triangle, $\gamma=3\beta$; in the leptonic triangle, $|\alpha_\ell|=2|\gamma_\ell|$ in the upper octant and $|\gamma_\ell|=2|\alpha_\ell|$ in the lower, the ratios visible in Fig.~\ref{fig:pmnstriangle}. The proof is tick arithmetic. The angle sum is $24$ ticks, and the maximal-phase image sets $\alpha=12$, so $\beta+\gamma=12$; with $\beta=3$ from the theorem, $\gamma=9=3\beta$. For the leptons, a $\beta$-preserving transfer of $t$ ticks carries $(12,3,9)$ to $(12+t,\,3,\,9-t)$, and the doubling $|\alpha_\ell|=2|\gamma_\ell|$ holds if and only if $12+t=2(9-t)$, that is, $t=2$; the two-unit transfer of Eq.~(\ref{eq:transfer}) and the doubling are therefore equivalent statements, and the octant exchange, which interchanges $|\alpha_\ell|$ and $|\gamma_\ell|$, carries the relation to $|\gamma_\ell|=2|\alpha_\ell|$.

The theorem and its corollaries thus predict the complete angle content of every triangle in the analysis. On the clock, the quark triangle is fixed outright, $(\alpha,\beta,\gamma)=(90^\circ,22.5^\circ,67.5^\circ)$, and the leptonic triangle is fixed as the unordered set $\{105^\circ,\,52.5^\circ,\,22.5^\circ\}$ in either octant, with only the assignment of the two large angles to their vertices ambiguous, the exchange the octant resolves. Up to that single twofold triangle ambiguity, all six angles of the two sectors are outputs of the postulate.

\section{The leptonic \texorpdfstring{$\pi/24$}{pi/24} triangle}
\label{sec:pmns}

The lepton sector inverts the quark-sector situation twice over. Its mixing angles are large, so by the density argument of Sec.~\ref{sec:ninths} the multiplicative lattice has no purchase on them; only the angular structure can act. And its data are in the opposite state of completion; the mixing angles are known to the percent level \cite{NuFIT2024}, while the CP phase carries an uncertainty of several tens of degrees and the atmospheric octant is unresolved. The angular hypothesis therefore changes character; what was a consistency test for the quarks becomes a prediction for the leptons. Throughout, the PMNS matrix is taken to be unitary, as in the global fits from which the inputs are drawn; the relations stated here are relations among the parameters of that unitary description and are not tests of unitarity itself.

The PMNS analog of the $B$-physics triangle is the $(1,3)$-column relation
\begin{equation}
U_{e1}U_{e3}^*+U_{\mu1}U_{\mu3}^*+U_{\tau1}U_{\tau3}^*=0,
\end{equation}
with angles defined in parallel to $(\alpha,\beta,\gamma)$,
\begin{equation}
\begin{gathered}
\alpha_\ell=\arg\!\left(-\frac{U_{\tau1}U_{\tau3}^*}{U_{e1}U_{e3}^*}\right),\qquad
\beta_\ell=\arg\!\left(-\frac{U_{\mu1}U_{\mu3}^*}{U_{\tau1}U_{\tau3}^*}\right),\\
\gamma_\ell=\arg\!\left(-\frac{U_{e1}U_{e3}^*}{U_{\mu1}U_{\mu3}^*}\right).
\end{gathered}
\end{equation}
Majorana phases multiply columns of $U$ and cancel in each ratio, so this triangle is a pure Dirac observable, and unlike its quark counterpart it is far from degenerate; all three angles are large. We place this triangle, rather than any of the other five nondegenerate PMNS triangles, because it is the direct analog of the CKM $(1,3)$-column ($db$) triangle used above, permitting the like-for-like comparison of Eq.~(\ref{eq:transfer}); the remaining triangles have not been scanned for lattice placements, and no selection credit is claimed.

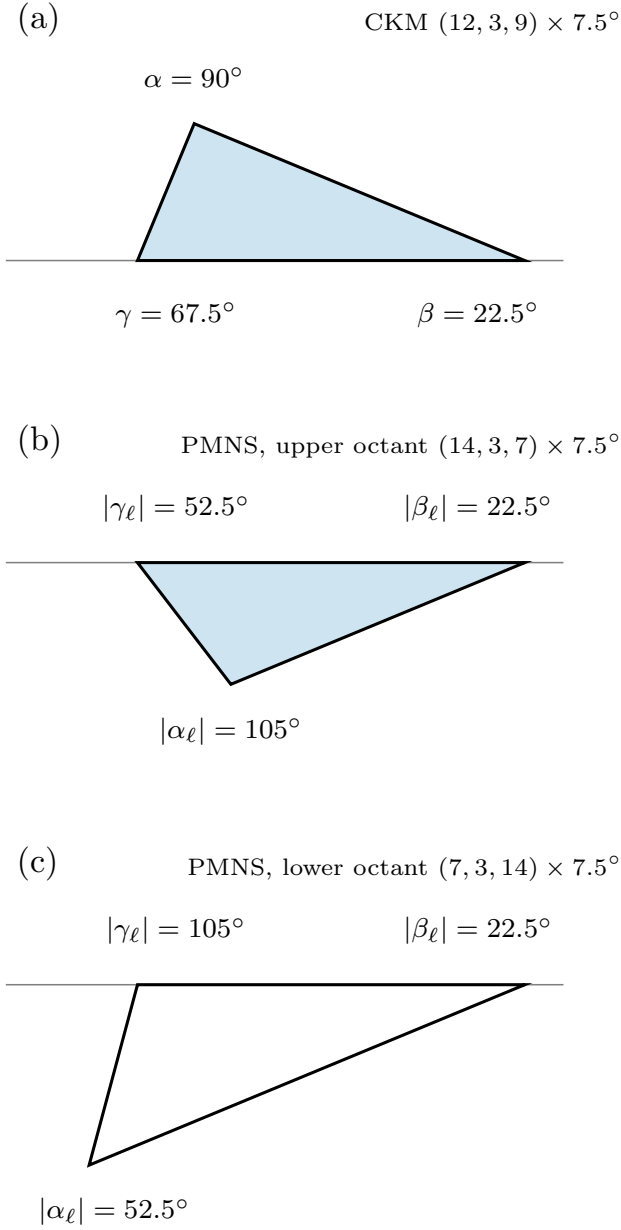
\begin{figure}[t]
\centering
\resizebox{0.96\columnwidth}{!}{%
\begin{tikzpicture}[x=4.0cm,y=4.0cm]
\begin{scope}[shift={(0,0.00000)}]
\fill[cband,opacity=0.5] (0,0) -- (0.14645,0.35355) -- (1,0) -- cycle;
\draw[black!50,line width=0.45pt] (-0.34,0) -- (1.10,0);
\draw[black,line width=0.9pt] (0,0) -- (0.14645,0.35355) -- (1,0) -- cycle;
\node[font=\normalsize,anchor=south west] at (-0.34,0.55000) {(a)};
\node[font=\scriptsize,anchor=south east] at (1.28,0.55000) {CKM $(12,3,9)\times7.5^\circ$};
\node[font=\footnotesize] at (0.88,-0.14000) {$\beta=22.5^\circ$};
\node[font=\footnotesize] at (0.10,-0.14000) {$\gamma=67.5^\circ$};
\node[font=\footnotesize] at (0.14645,0.47355) {$\alpha=90^\circ$};
\end{scope}
\begin{scope}[shift={(0,-0.78000)}]
\fill[cband,opacity=0.5] (0,0) -- (0.24118,-0.31431) -- (1,0) -- cycle;
\draw[black!50,line width=0.45pt] (-0.34,0) -- (1.10,0);
\draw[black,line width=0.9pt] (0,0) -- (0.24118,-0.31431) -- (1,0) -- cycle;
\node[font=\normalsize,anchor=south west] at (-0.34,0.24000) {(b)};
\node[font=\scriptsize,anchor=south east] at (1.28,0.24000) {PMNS, upper octant $(14,3,7)\times7.5^\circ$};
\node[font=\footnotesize] at (0.88,0.14000) {$|\beta_\ell|=22.5^\circ$};
\node[font=\footnotesize] at (0.10,0.14000) {$|\gamma_\ell|=52.5^\circ$};
\node[font=\footnotesize] at (0.24118,-0.43431) {$|\alpha_\ell|=105^\circ$};
\end{scope}
\begin{scope}[shift={(0,-1.87000)}]
\draw[black!50,line width=0.45pt] (-0.34,0) -- (1.10,0);
\draw[black,line width=0.9pt] (0,0) -- (-0.12484,-0.46593) -- (1,0) -- cycle;
\node[font=\normalsize,anchor=south west] at (-0.34,0.24000) {(c)};
\node[font=\scriptsize,anchor=south east] at (1.28,0.24000) {PMNS, lower octant $(7,3,14)\times7.5^\circ$};
\node[font=\footnotesize] at (0.88,0.14000) {$|\beta_\ell|=22.5^\circ$};
\node[font=\footnotesize] at (0.10,0.14000) {$|\gamma_\ell|=105^\circ$};
\node[font=\footnotesize] at (-0.06000,-0.58593) {$|\alpha_\ell|=52.5^\circ$};
\end{scope}
\end{tikzpicture}}
\caption{The $\pi/24$ unitarity triangles, normalized to unit base and stacked at a common scale: (a)~the CKM triangle; (b)~the PMNS $(1,3)$-column triangle in the upper octant and (c)~in the lower octant, with $\delta_{\rm CP}$ at the values of Eq.~(\ref{eq:pmnspred}); the octant exchange reflects the apex through the mid-perpendicular of the base. Panels (a) and (b), the pair linked by the transfer relation $\delta_{\rm CP}\simeq-\delta$ (opposite Dirac phases in the two sectors), are shaded; (c), the configuration that relation excludes, is not. The PMNS apexes lie below their base and the CKM apex above, displaying ${\rm sign}(J_\ell)=-\,{\rm sign}(J)$, and the angle at the right-hand vertex is $22.5^\circ$ in all three panels, the shared $\beta=\beta_\ell$. Panel (b) is the best-fit branch, the filled blue star of Fig.~\ref{fig:stake}; panel (c) is its open-star alternative.}
\label{fig:pmnstriangle}
\end{figure}

\begin{figure*}[t]
\centering
\resizebox{0.70\textwidth}{!}{%
\begin{tikzpicture}
\begin{axis}[width=13.2cm,height=9.2cm,
  xmin=170,xmax=370,ymin=0.375,ymax=0.66,
  xlabel={$\delta_{\rm CP}$ [deg]},ylabel={$\sin^2\theta_{23}$},
  tick label style={font=\normalsize},axis line style={line width=0.5pt}]
\fill[cband,opacity=0.6] (axis cs:293,0.375) rectangle (axis cs:296,0.66);
\draw[black!70,line width=0.8pt,dotted] (axis cs:170,0.5) -- (axis cs:370,0.5);
\draw[black!70,line width=0.8pt,dashed] (axis cs:180,0.375) -- (axis cs:180,0.66);
\draw[black!70,line width=0.8pt,dashed] (axis cs:360,0.375) -- (axis cs:360,0.66);
\draw[black!70,line width=0.8pt,dashdotted] (axis cs:270,0.375) -- (axis cs:270,0.66);
\node[rotate=90,anchor=south west,font=\small] at (axis cs:264.5,0.505) {maximal CP violation};
\node[rotate=90,anchor=south west,font=\small] at (axis cs:183,0.392) {CP conserving};
\node[rotate=90,anchor=south west,font=\small] at (axis cs:352.8,0.392) {CP conserving};
\node[anchor=south,font=\small] at (axis cs:342,0.503) {maximal $\theta_{23}$};
\node[font=\small,text=cblue] at (axis cs:294.5,0.640) {$360^\circ-\delta=294.5^\circ\pm1.5^\circ$};
\node[star,star points=5,star point ratio=2.35,minimum size=13pt,inner sep=0pt,
  draw=cblue,fill=cblue] at (axis cs:296.2,0.549) {};
\node[star,star points=5,star point ratio=2.35,minimum size=13pt,inner sep=0pt,
  draw=cred,fill=white,line width=0.8pt] at (axis cs:243.8,0.451) {};
\node[font=\small,text=cblue] at (axis cs:296.2,0.578) {NO, IC19 w/o SK-atm};
\node[font=\small,text=cred] at (axis cs:243.8,0.420) {NO, IC24 with SK-atm};
\draw[line width=0.7pt,{Bar[width=5pt]}-{Bar[width=5pt]}] (axis cs:281,0.408) -- (axis cs:311,0.408);
\node[font=\small] at (axis cs:296,0.394) {anticipated $\pm15^\circ$};
\end{axis}
\end{tikzpicture}}
\caption{The octant stake. Stars are the two full-triple tangency solutions, $(\sin^2\theta_{23},\delta_{\rm CP})=(0.549,\,296.2^\circ)$ and $(0.451,\,243.8^\circ)$, each labeled by the NuFit-6.0 analysis variant \cite{NuFIT2024} whose normal-ordering best fit selects that octant (IC19 without SK-atm: upper; IC24 with SK-atm: lower); the filled blue star is the framework's best-fit solution, the upper-octant branch selected by the transfer relation of Eq.~(\ref{eq:transfer}), and the open red star is the lower-octant alternative; the vertical band is $\delta_{\rm CP}=360^\circ-\delta=294.5^\circ\pm1.5^\circ$ from the measured CKM phase $\delta$, coinciding with the upper-octant solution only. The dotted line is maximal $\theta_{23}$, dashed lines are the CP-conserving points, the dash-dotted line marks maximal CP violation, $\delta_{\rm CP}=270^\circ$, and the error bar illustrates the anticipated DUNE and Hyper-Kamiokande resolution on $\delta_{\rm CP}$.}
\label{fig:stake}
\end{figure*}
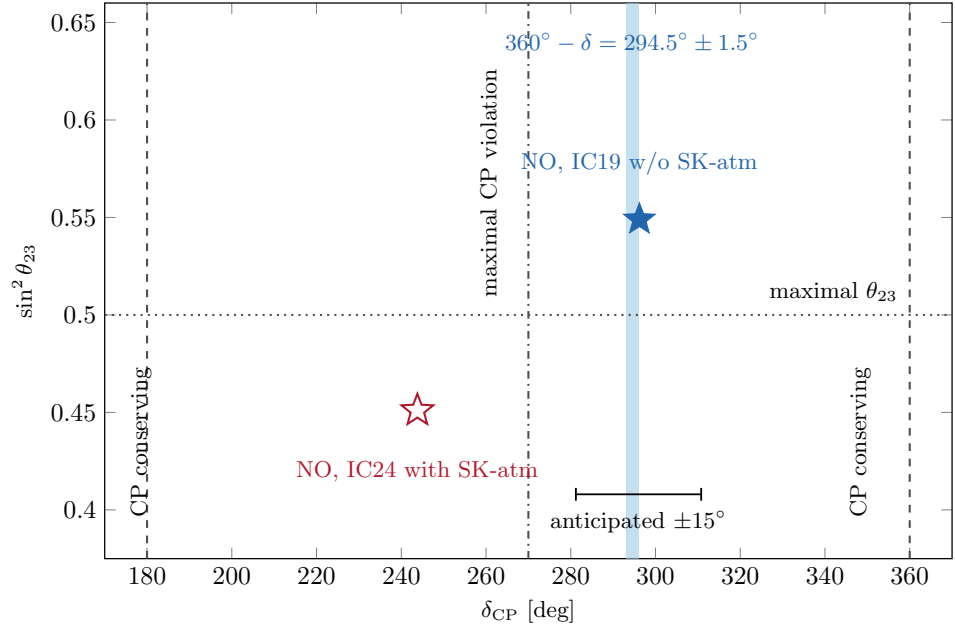

Using $\sin^2\theta_{12}=0.307$ and $\sin^2\theta_{13}=0.0220$ \cite{NuFIT2024}, with benchmark atmospheric values $\sin^2\theta_{23}=0.553$ (upper octant) and $0.455$ (lower octant), within $0.5\sigma$ and $1.2\sigma$ of the corresponding variant best fits, $0.561$ (IC19 w/o SK-atm) and $0.470$ (IC24 with SK-atm) \cite{NuFIT2024}, the nondegenerate $\pi/24$ solutions in the CP-violating half-plane $\delta_{\rm CP}\in(180^\circ,360^\circ)$ are
\begin{equation}
\begin{split}
\text{upper octant:}\;\;
(|\alpha_\ell|,|\beta_\ell|,|\gamma_\ell|)
&=(14,3,7)\times7.5^\circ\\
&=(105^\circ,\,22.5^\circ,\,52.5^\circ),\\
\delta_{\rm CP}&\simeq296^\circ,\\
\text{lower octant:}\;\;
(|\alpha_\ell|,|\beta_\ell|,|\gamma_\ell|)
&=(7,3,14)\times7.5^\circ\\
&=(52.5^\circ,\,22.5^\circ,\,105^\circ),\\
\delta_{\rm CP}&\simeq244^\circ.
\end{split}
\label{eq:pmnspred}
\end{equation}
Mirror solutions at $360^\circ-\delta_{\rm CP}$ flip the sign of the leptonic Jarlskog invariant, and the octant exchange $\theta_{23}\to90^\circ-\theta_{23}$ interchanges $|\alpha_\ell|\leftrightarrow|\gamma_\ell|$. Under the exchange the solved phase maps exactly to $540^\circ-\delta_{\rm CP}$, a reflection through $270^\circ$, so the two octant solutions sit symmetrically about maximal CP violation, at $270^\circ\pm26^\circ$, as Fig.~\ref{fig:stake} displays. The predictions are insensitive to the octant benchmarks; varying $\sin^2\theta_{23}$ by $\pm0.02$ about either input moves the solved $\delta_{\rm CP}$ by less than $0.5^\circ$, so the role of $\theta_{23}$ is to select the branch, not to set the value; Fig.~\ref{fig:pmnstriangle} draws both octant triangles against the quark triangle, with the shared $\beta=22.5^\circ$ and the opposite CP orientations visible directly.

Read at full strength, the clock hypothesis determines $\theta_{23}$ as well. Along the curve of solutions with $\beta_\ell=22.5^\circ$, the angle $|\alpha_\ell|$ does not cross $105^\circ$ but touches it, reaching a minimum of $105.0^\circ$; the complete triple $(105^\circ,22.5^\circ,52.5^\circ)$ is therefore attained tangentially, at one point per octant, $\sin^2\theta_{23}=0.549$ with $\delta_{\rm CP}=296.2^\circ$ in the upper octant and, by the exact exchange, $0.451$ with $243.8^\circ$ in the lower. Demanding the full triple thus predicts the deviation from maximal atmospheric mixing, $|\sin^2\theta_{23}-\tfrac12|\simeq0.05$, against the variant best fits $0.561$ and $0.470$ \cite{NuFIT2024} at $0.8\sigma$ and $1.1\sigma$; exact maximality misses the triple by $0.6^\circ$ in $|\alpha_\ell|$ and is mildly disfavored. The tangency also explains the insensitivity noted above; the triple sits at an extremum along the solution curve, so departures are quadratic. Two cautions accompany this reading. The tangency location is stable, $0.549\pm0.007$ under $1\sigma$ variations of the solar inputs, but its depth is not; $|\alpha_\ell|_{\rm min}$ moves from $101.7^\circ$ to $108.0^\circ$ across the $\theta_{12}$ uncertainty, so the near-exact touch at central inputs is a $0.3\sigma$ statement, one further relation among the measured angles that enters the same look-elsewhere accounting as the rest, counted and not compounded.
 Within the current $1\sigma$ ranges of $\theta_{12}$ and $\theta_{13}$ the quoted $\delta_{\rm CP}$ values shift at the level of a few degrees, and for excursions at the edges of those ranges distinct lattice configurations can become competitive; Eq.~(\ref{eq:pmnspred}) should be read as a central-value statement, to be refit as the angles sharpen.

On the $\pi/24$ hypothesis both triangles share $\beta=\beta_\ell=22.5^\circ$, three half-units; on the quark side this is the best-measured angle and agrees at $0.4\sigma$, while on the lepton side it is a prediction.

The two triangles are related more closely still. Their lattice assignments differ by a transfer of exactly two half-units from $\gamma$ to $\alpha$,
\begin{equation}
\begin{split}
(|\alpha_\ell|,|\beta_\ell|,|\gamma_\ell|)-(\alpha,\beta,\gamma)
&\;=\;(14,3,7)-(12,3,9)\\
&\;=\;(+2,\,0,\,-2),
\end{split}
\label{eq:transfer}
\end{equation}
so that $\alpha_\ell-\alpha=\gamma-\gamma_\ell=15^\circ$ in the upper octant. At the level of the PDG phases this is numerically the statement that the leptonic and quark Dirac phases are equal and opposite, $\delta_{\rm CP}\simeq-\delta$ (mod $360^\circ$);\footnote{In sector-labeled notation the relation reads $\delta_{\rm CP}\simeq-\delta_{\rm CKM}$, with $\delta_{\rm CKM}\equiv\delta$. We retain the standard symbols throughout: bare $\delta$ for the CKM Dirac phase, as in the PDG parameterization, and $\delta_{\rm CP}$ for the leptonic phase, since that is the symbol in which DUNE and Hyper-Kamiokande will report.} with the measured CKM phase $\delta=65.5^\circ\pm1.5^\circ$, this places the leptonic phase at $360^\circ-\delta=294.5^\circ\pm1.5^\circ$, within $1.7^\circ$ of the lattice value $296.2^\circ$, so the two formulations are indistinguishable at any foreseeable precision and either way place the upper-octant target near $295^\circ$. A quantized leptonic phase has also been proposed in the conformal flavor-spin framework, $\delta_{\rm CP}=(19/12)\pi=285^\circ$ \cite{AigleJourjine2026}, which sits on the $\pi/24$ lattice at $38$ units, between the two octant solutions here; the same value is the inverted-ordering best-fit phase of the global fit, $285^{+25}_{-28}$ degrees in the IC19 variant \cite{NuFIT2024}, an ordering excluded in the present framework by the mass relation of Eq.~(\ref{eq:m23}).

Two qualifications accompany the relation. Stated through the phases it is convention-tied, since the PDG parameterization is used on both sides; the invariant content is Eq.~(\ref{eq:transfer}) together with ${\rm sign}(J_\ell)=-\,{\rm sign}(J)$, opposite senses of CP violation in the two sectors, with the magnitude ratio $|J_\ell/J|\approx970$ carried entirely by the angle prefactors. And given the shared $\beta$ and triangle closure, Eq.~(\ref{eq:transfer}) and $\delta_{\rm CP}\simeq-\delta$ are one relation, not two.

The relation also constitutes an octant prediction, and we place the stake explicitly. The value $360^\circ-\delta=294.5^\circ\pm1.5^\circ$ coincides with the upper-octant solution of Eq.~(\ref{eq:pmnspred}) and misses the lower-octant solution by $51^\circ$, so Eq.~(\ref{eq:transfer}) can hold only for $\theta_{23}>45^\circ$; a resolution of the octant in favor of $\theta_{23}<45^\circ$ falsifies the transfer relation outright, independently of any $\delta_{\rm CP}$ measurement, leaving only the mirror lattice solution at $244^\circ$. The coherent corner of parameter space for the combined structure is therefore normal ordering, upper octant, and $\delta_{\rm CP}\simeq295^\circ$. Present data neither select nor exclude it; both lattice values lie inside the $3\sigma$ normal-ordering $\delta_{\rm CP}$ ranges of the two analysis variants while outside $1\sigma$, against best-fit phases of $177^{+19}_{-20}$ degrees (IC19 w/o SK-atm) and $212^{+26}_{-41}$ degrees (IC24 with SK-atm) \cite{NuFIT2024}, and the octant preference of current global fits changes sign between analysis variants \cite{NuFIT2024}, the IC19 fit without SK atmospheric data preferring the upper octant and the IC24 fit including the SK-atm samples the lower, the robust content of the $\mu$--$\tau$ block being the magnitude of the deviation from maximality rather than its sign, and the appearance channels that carry the octant information are entangled with $\delta_{\rm CP}$ itself. The variant split reflects the treatment of the externally supplied SK-atm $\chi^2$ table rather than a ranking; the same sample that drives the lower-octant preference also supplies the normal-ordering preference, $\Delta\chi^2=6.1$, without which the two orderings are essentially degenerate (normal ordering at $\Delta\chi^2=0.6$) \cite{NuFIT2024}, so the coherent corner draws support from both variants and coincides with neither. The upper-octant configuration of Fig.~\ref{fig:pmnstriangle}(b) is thus the framework's prediction, distinguished by a relation falsifiable through the octant alone, and DUNE and Hyper-Kamiokande test it by two independent routes on one schedule.

The phase route confronts the measured $\delta_{\rm CP}$, extracted from the neutrino--antineutrino appearance asymmetry, with $296^\circ$ at the anticipated $\pm15^\circ$ precision. The octant route uses the appearance-rate normalization, which determines $\sin^2\theta_{23}$ itself rather than the octant-blind $\sin^22\theta_{23}$ of the disappearance channel, and falsifies the relation outright if $\theta_{23}<45^\circ$, whatever value of $\delta_{\rm CP}$ is found. Figure~\ref{fig:stake} displays the stake.

The survival map is worth stating in one place. The leptonic quantization itself lives only if the measured pair $(\delta_{\rm CP},\,\mathrm{octant})$ lands on one of the two starred solutions of Fig.~\ref{fig:stake}; the transfer relation, and with it the coherent corner of the combined structure, lives only on the blue one; and normal ordering is prerequisite to both, through Eq.~(\ref{eq:m23}). A measured phase near the current variant best fits, $177^\circ$ or $212^\circ$, would exclude the blue solution outright and leave the red one strained at the $2\sigma$ level or beyond at the anticipated precision.

The first row of the PMNS matrix admits a mass-ratio description of its own. Writing $r\equiv m_2/m_3=0.1719\pm0.0026$, the measured moduli satisfy
\begin{equation}
\big(|U_{e1}|,\,|U_{e2}|,\,|U_{e3}|\big)\;\simeq\;\big(r^{1/9},\;r^{3/9},\;\tfrac{\sqrt{3}}{2}\,r\big)
\label{eq:firstrow}
\end{equation}
with coefficients $(1.001,\,0.985,\,0.996)$; unitarity makes these two independent relations rather than three, and the statement deserves its quantitative form.

Taken all three as exact, the displayed powers over-close the row, $\sum_i|U_{ei}|^2=1.0076$, with the excess concentrated in the $e2$ entry. The exact statements are the angle-level relations, $\sin\theta_{13}=\tfrac{\sqrt{3}}{2}\sqrt{\Delta m^2_{21}/\Delta m^2_{31}}$, agreeing at $0.19\sigma$, and $\sin\theta_{12}=(m_2/m_3)^{1/3}$, equivalently $\sin^6\theta_{12}=\Delta m^2_{21}/\Delta m^2_{31}$, agreeing at $0.18\sigma$; the row then follows unitarily through the standard $c_{13}$ factors, each entry within $0.4\%$ of data (a first-row near identity has also been reported in the conformal flavor-spin framework \cite{AigleJourjine2026}), and Eq.~(\ref{eq:firstrow}) is the leading form of that unitary row, accurate up to $c_{13}=1-\tfrac{3}{8}r^2+\ldots$, a $1.1\%$ correction. The undressed relation $\sin\theta_{13}\simeq\sqrt{\Delta m^2_{21}/\Delta m^2_{31}}$ is now excluded at high significance; the $\sqrt{3}/2$ dressing is what current data select. The $\theta_{13}$ relation chains onto the eighteenths lattice through Eq.~(\ref{eq:m23}), $\sin\theta_{13}=\tfrac{\sqrt{3}}{2}\,\e^{19/18}$; the $\theta_{12}$ relation does not, since $r^{1/3}=\e^{19/54}$ breaks the eighteenths denominator, so the row package is a distinct structure, powers of the measured ratio $r$ itself with exponents tripling as $(1,3,9)$ in ninths.

Because these are order-one relations with fixed constants, the family of candidate constants and small rational exponents implies a look-elsewhere effect we cannot rigorously bound, and we do not claim a chance probability below the tens-of-percent level for the package as retrodiction; its weight rests on the near-term tests, since JUNO will determine $\Delta m^2_{21}/\Delta m^2_{31}$ at the few-per-mille level and sharpen $\theta_{12}$ substantially, turning both relations into sub-percent confrontations.

The lightest mass completes the spectrum, and we record the lattice-consistent completion as a hypothesis; it is not a fit, and no chance-probability credit attaches to it. Repeating the measured step gives the geometric spectrum $(m_1,m_2,m_3)=m_3\,(\varrho^2,\varrho,1)$ with $\varrho=m_2/m_3=\e^{19/18}$, hence
\begin{equation}
m_1=\frac{m_2^2}{m_3}=1.48\pm0.02~{\rm meV}\,;
\label{eq:m1}
\end{equation}
the unrefined alternative $m_1=\e^2 m_3=1.76$~meV is observationally indistinguishable from it.

Under the geometric completion the first-row package of Eq.~(\ref{eq:firstrow}) takes a unified form,
\begin{equation}
\begin{split}
\big(|U_{e1}|,\,|U_{e2}|,\,|U_{e3}|\big)
\;=\;\Big(&\big(\tfrac{m_1}{m_3}\big)^{1/18},\;\big(\tfrac{m_1}{m_3}\big)^{3/18},\\
&\;\;\tfrac{\sqrt{3}}{2}\big(\tfrac{m_1}{m_3}\big)^{9/18}\Big),
\end{split}
\label{eq:unifiedrow}
\end{equation}
with the tripling exponents $(1,3,9)$ now over the denominator eighteen; conditional on Eq.~(\ref{eq:m1}), the same denominator that organizes the quark lattice appears in the lepton first row through the extreme mass ratio. As with Eq.~(\ref{eq:firstrow}), this is the leading form; the exact unitary version assigns the powers to $\sin\theta_{12}=(m_1/m_3)^{1/6}$ and $\sin\theta_{13}=\tfrac{\sqrt{3}}{2}(m_1/m_3)^{1/2}$, with the row completed by the standard $c_{13}$ factors.

The observable corollaries are a summed mass $\Sigma m_\nu=60.6$~meV, against $59.0$~meV for a massless lightest state, for the adopted splittings $\Delta m^2_{21}=7.49\times10^{-5}~{\rm eV}^2$ and $\Delta m^2_{31}=2.534\times10^{-3}~{\rm eV}^2$ \cite{NuFIT2024}, squarely in the minimal normal-ordering band that upcoming cosmological measurements test, and an effective Majorana mass confined to $m_{\beta\beta}\in[0.5,\,4.7]$~meV over the Majorana phases, below the sensitivity of the ton-scale $0\nu\beta\beta$ program; within this spectrum a positive ton-scale signal would be incompatible with the completion. The completion also acts as a discriminant among texture proposals; the class predicting $\tan^2\theta_{12}=m_1/m_2$ requires $0.44$ where the geometric spectrum gives $0.17$, and is excluded within the framework.

The charged-lepton masses, absent from the mixing moduli and from the lattice inventory of Table~\ref{tab:inventory}, enter this structure at exactly one point. The measured splitting ratio satisfies
\begin{equation}
\frac{\Delta m^2_{21}}{\Delta m^2_{31}} \;=\; \frac{m_\mu}{2\,m_\tau}\,,
\label{eq:massmass}
\end{equation}
with the $M_Z$ (pole) mass ratio giving $0.02941$ ($0.02973$) against the measured $0.02956\pm0.00090$, agreement at the $0.2\sigma$ level at either scale; chained with Eq.~(\ref{eq:firstrow}) it gives $\sin^2\theta_{13}=\tfrac{3}{8}\,m_\mu/m_\tau=0.0221$ against $0.0220\pm0.0006$. Only one of these three relations is independent, and the same look-elsewhere caution applies to the factor of two. What distinguishes Eq.~(\ref{eq:massmass}) is its test; $m_\mu/m_\tau$ is known to the $10^{-4}$ level, so the relation is a fixed-number prediction for the splitting ratio, $0.0294$--$0.0297$ with the spread set by the renormalization scale of the charged ratio, and JUNO's few-per-mille determination will land inside or outside that band. Geometric-mean forms of the charged masses, such as $\sqrt{m_e m_\tau}/m_\mu$ and quarter-root ratios, match no off-diagonal PMNS modulus within $3\sigma$; the charged sector ties to the neutrino mass spectrum, not to the mixing.

The second and third rows show no comparable placement. Their moduli depend on $\theta_{23}$ and $\delta_{\rm CP}$, the quantities the division of labor of Sec.~\ref{sec:discussion} assigns to the angular structure, and at the benchmarks of Eq.~(\ref{eq:pmnspred}) their fitted exponents in ninths of $r$ scatter with mean integer distance $0.26$ (upper octant) and $0.23$ (lower), against a chance expectation of $0.25$. Any apparent placement among them is also degenerate with the unmeasured $\delta_{\rm CP}$, which can park individual $\mu$- or $\tau$-row moduli on so coarse a lattice by choice; we therefore attach no weight to such placements. The boundary between the mass-ratio and angular descriptions thus runs between the first row and the $\mu$--$\tau$ block; the first row follows from the mass spectrum, while the remaining parameters, near-maximal $\theta_{23}$ and the lattice value of $\delta_{\rm CP}$, are angular statements.

An experimental resolution of $\delta_{\rm CP}$ at the level of $10^\circ$--$20^\circ$, as anticipated from DUNE and Hyper-Kamiokande over favorable regions of parameter space, tests Eq.~(\ref{eq:pmnspred}) against the CP-conserving points and against generic values, provided the octant is resolved. Because these are predictions rather than accommodations, no chance-probability discount of the kind computed in Sec.~\ref{sec:stats} applies to them; they will be right or wrong.

\section{Discussion}
\label{sec:discussion}

Three points situate this analysis.

First, its epistemic status. Every statement in this paper is a statement about measured numbers; the FX inversion is exact, the lattice assignments are fits with quantified integer distances and coefficients, the triangle patterns are compared with rephasing invariants, and the chance probabilities are computed and reported rather than implied. Nothing here requires, presumes, or gestures at an ultraviolet completion, a symmetry, or a mechanism, and the analysis is not vulnerable to the absence of one, any more than the Wolfenstein expansion was \cite{Wolfenstein1983}. The tradition of parameterization-level regularities \cite{GST1968,HPS2002,Minakata2004,Raidal2004} shows both the utility and the correct posture of such work; the regularity is stated, its precision and its chance probability are quantified, and the field decides its meaning through the tests it enables.

Second, the division of labor between the two structures, which we regard as the central lesson. The multiplicative lattice is a hierarchy statement and is empty for order-one quantities; the angular lattice quantizes the triangle angles in units of $\pi/24$ and, as a rephasing-invariant statement, applies wherever a nondegenerate triangle exists. Quark mixing is hierarchical and realizes both, with the angular pattern arising as the image of the multiplicative one at maximal phase through $\alpha\simeq\phifx$ and $\tan\beta\simeq\sqrt{\e}$. Lepton mixing is large, admits only the angular structure, and converts it into a prediction for $\delta_{\rm CP}$. The two sectors are thus not forced into a single template; they realize complementary faces of one commensurate description, meeting in unitarity triangles that share $\beta=22.5^\circ$, the one numerical statement the two sectors hold in common. Together the two structures render flavor digital; the data carry integer exponents multiplicatively and integer ticks angularly, with the order-one coefficients and the residuals quantified above as the measure of how far that statement holds.

Third, the requirements on theory. Any dynamical account of these patterns would need to explain five things: why the FX angles carry integer exponents on a lattice of eighteenths, the signature pattern of Froggatt--Nielsen charge counting \cite{FN1979}; why the single CP phase is maximal; why the anchor scale satisfies $\e\simeq0.187$, within ten percent of, but distinct from, the angular fixed point $3-2\sqrt{2}$; why the same $\e$ that sets the CKM hierarchy also sets the neutrino mass ratio through $m_2/m_3=\e^{19/18}$, with the PMNS first row then expressible in powers of that ratio with tripling exponents and the charged-lepton masses entering only through $\Delta m^2_{21}/\Delta m^2_{31}=m_\mu/(2m_\tau)$, while the charged-lepton ratios at $M_Z$ sit off the lattice; and why the leptonic triangle, if the $\delta_{\rm CP}$ prediction succeeds, lands on the same $\pi/24$ lattice from an entirely different mixing regime, displaced from the quark triangle by the two-unit transfer of Eq.~(\ref{eq:transfer}) with opposite-sign CP violation.

Of these, the maximal phase has a known candidate answer at the symmetry level; an antiunitary $Z_2$, an exchange folded with complex conjugation, quantizes the phase it acts on to $\pm\pi/2$ with no intermediate value, the folding argument developed for the leptonic sector in the $\pi/12$ companion \cite{Pi12Model}, where the unitary half of the same reflection pins $\theta_{23}=\pi/4$ on the mirror bisector. A maximal $\phifx$ would then read as the trace of such a folding, and the digital clock's right angle, $\alpha\simeq\phifx$ at twelve ticks, as its image on the dial. We otherwise take no position on candidate mechanisms; the purpose of an empirical benchmark is to make such questions precise, and the measurements that will adjudicate the benchmarks, degree-level $\gamma$ from LHCb and Belle~II and $\delta_{\rm CP}$ from DUNE and Hyper-Kamiokande, are scheduled rather than hypothetical.

\section{Summary}
\label{sec:summary}

The measured FX parameters of the CKM matrix realize two commensurate structures at once. Multiplicatively, with the hierarchy parameter defined internally as $\e=|V_{ub}|^{3/10}=0.1869\pm0.0018$, the three angles sit on a lattice of eighteenths, $\sin\theta_u=\e^{26/18}$, $\sin\theta_d=\e^{17/18}$, $\sin\theta=\e^{34/18}$, with coefficients within $2.5\%$ of unity, and the phase sits at its maximal value, $\phifx=92.3^\circ\pm2.7^\circ$ against $\pi/2$; with unit coefficients this reproduces the remaining moduli at the percent level and fixes $J=\e^{111/18}\sin\phifx$; at the element level, all six off-diagonal moduli are generated by four exponents, $(16,34,51,60)$, built from the three angle integers, with the anchor choice between $|V_{ub}|^{3/10}$ and $|V_{us}|^{9/8}$ immaterial at the $0.1\%$ level; in base $\lambda=\Vus$ the lattice reads as a fractional-power Wolfenstein expansion with $A=\lambda^{1/8}$ and $|1-\bar\rho-i\bar\eta|=\lambda^{1/16}$.

The quark mass ratios at $M_Z$ join the same lattice, $(m_d/m_s,\,m_s/m_b,\,m_c/m_t)=\e^{(32,43,60)/18}$, yielding the parameter-free relations $m_d/m_s=\lambda^2$, $m_s/m_b=\sin\theta_d\sin\theta_u$ (exact to $0.02\sigma$), and $m_c/m_t=|V_{ub}|$, and with them a construction of the full CKM matrix from three mass ratios and the maximal phase, in which the up-sector ratio $m_u/m_c$ and its factor-of-two failure play no part.

Additively, the triangle angles are quantized in units of $\pi/24=7.5^\circ$; the rephasing-invariant image is the unitarity triangle $(\alpha,\beta,\gamma)=(12,3,9)\times7.5^\circ=(90^\circ,22.5^\circ,67.5^\circ)$, consistent with all current data at the $1\sigma$ level through the map $\alpha\simeq\phifx$ and $\tan\beta\simeq\sqrt{\e}$; exact coincidence of the two lattices would require $\e=3-2\sqrt{2}$, which the moduli exclude, so the additive pattern is read as the angular approximation of the multiplicative one at maximal phase. The map admits a theorem; $\tan\beta=\sqrt{\e}$, giving $\beta=23.0^\circ$ at unit coefficients, with corollaries $\gamma=3\beta$ for the quarks and the doubling $|\alpha_\ell|=2|\gamma_\ell|$ (upper octant) or $|\gamma_\ell|=2|\alpha_\ell|$ (lower) for the leptons, the latter equivalent to the two-unit transfer, so that all six triangle angles are outputs up to the octant exchange.

The chance probabilities of the retrodictive agreements are $0.10$ and $0.04$ and are reported as such; the case rests on the prospective tests, degree-level $\gamma$ against $67.5^\circ$--$68^\circ$ on the quark side and, on the lepton side, where only the angular structure can act, the prediction $\delta_{\rm CP}\simeq296^\circ$ (upper octant) or $244^\circ$ (lower) from the Majorana-phase-free $(1,3)$-column PMNS triangle, quantized on the same lattice at $(14,3,7)$ or $(7,3,14)\times7.5^\circ$, with $\beta_\ell=22.5^\circ$ shared between the quark and lepton triangles; read at full strength the triple also fixes the atmospheric deviation, $\sin^2\theta_{23}=0.549$ (upper) or $0.451$ (lower), within $0.8\sigma$ and $1.1\sigma$ of the variant best fits.

Two cross-sector relations tie the sectors together. First, the two triangles share $\beta=\beta_\ell=22.5^\circ$, the one angle common to the quark and lepton sectors, measured at $0.4\sigma$ on the quark side and predicted on the lepton side, and differ by a transfer of two lattice units, $(14,3,7)-(12,3,9)=(+2,0,-2)$, the null middle entry being the shared angle; at the phase level this is the statement that the leptonic and quark Dirac phases are equal and opposite, $\delta_{\rm CP}\simeq-\delta$, which the measured $\delta=65.5^\circ\pm1.5^\circ$ places at $360^\circ-\delta=294.5^\circ\pm1.5^\circ$, within $1.7^\circ$ of the lattice value $296.2^\circ$ and indistinguishable from it at any foreseeable precision; the relation holds only in the upper atmospheric octant, so the octant resolution alone can falsify it. Second, the same $\e$ that organizes the CKM hierarchy sets the neutrino mass ratio, $m_2/m_3=\e^{19/18}$ with coefficient $1.010\pm0.018$, equivalently $|V_{ub}|=(m_2/m_3)^{60/19}$ at the $3\%$ level, while the charged-lepton mass ratios evaluated at $M_Z$ sit off the lattice and are excluded from the inventory.

The PMNS first row is expressible in powers of the same ratio, $(|U_{e1}|,|U_{e2}|,|U_{e3}|)\simeq(r^{1/9},\,r^{1/3},\,\tfrac{\sqrt{3}}{2}r)$, two independent relations agreeing at $0.2\sigma$ and testable at JUNO, while the second and third rows show no comparable placement; the boundary between the mass-ratio and angular descriptions runs between the first row and the $\mu$--$\tau$ block. The charged-lepton masses enter at a single point, $\Delta m^2_{21}/\Delta m^2_{31}=m_\mu/(2m_\tau)$, a fixed-number prediction that JUNO will test at the per-mille level; and the geometric completion of the spectrum, $m_1=m_2^2/m_3=1.5$~meV, recorded as a hypothesis, places the first row at exponents $(1,3,9)/18$ of $m_1/m_3$ and yields $\Sigma m_\nu\simeq60$~meV with $m_{\beta\beta}\in[0.5,4.7]$~meV.

The analysis is confined to the parameterization level throughout, in the tradition of the empirical regularities that have historically preceded and guided dynamical understanding of flavor.

\clearpage

\begin{acknowledgments}
V.B. gratefully acknowledges support from the U.S. Department of Energy, Office of Science, Office of High Energy Physics, under Award Number DE-SC0017647 and from the William F. Vilas Estate.
\end{acknowledgments}

\end{document}